\documentclass{elsart}

\usepackage{graphicx}
\usepackage[figuresright]{rotating}
\usepackage[small]{caption}
\DeclareMathAlphabet{\mathsf}{OT1}{cmss}{m}{n}
\SetMathAlphabet{\mathsf}{bold}{OT1}{cmss}{bx}{n}

\usepackage{amssymb}

\newlength{\dslashwidth}

\newcommand{\tb}{\ensuremath{\tan\beta}}

\newcommand{\bq}{\begin{equation}}
\newcommand{\eq}{\end{equation}}
\newcommand{\ba}{\begin{array}}
\newcommand{\ea}{\end{array}}
\newcommand{\bqa}{\begin{eqnarray}}
\newcommand{\eqa}{\end{eqnarray}}

\newcommand{\lnf}{{\ifmmode \Lambda^{(N_f)} \else $\Lambda^{(N_f)}$\fi}}
\newcommand{\ms}{{\ifmmode \overline{MS} \else $\overline{MS}$\fi}}
\newcommand{\dr}{{\ifmmode \overline{DR} \else $\overline{DR}$\fi}}
\newcommand{\lms}{{\ifmmode \Lambda^{(5)}_{\overline{MS}} \else $\Lambda^{(5)}_{\overline{MS}}$\fi}}
\newcommand{\lam}{{\ifmmode \Lambda \else $\Lambda$\fi}}
\newcommand{\gev}{{\ifmmode {\rm GeV} \else ${\rm GeV}$\fi}}
\newcommand{\gevc}{{\ifmmode {\rm GeV/c^2} \else ${\rm GeV/c^2}$\fi}}
\newcommand{\tev}{{\ifmmode {\rm TeV} \else ${\rm TeV}$\fi}}
\newcommand{\tevc}{{\ifmmode {\rm TeV/c^2} \else ${\rm TeV/c^2}$\fi}}
\newcommand{\lp}{{\ifmmode L^+  \else $L^+$\fi}}
\newcommand{\lm}{{\ifmmode L^-  \else $L^-$\fi}}
\newcommand{\mlp}{{\ifmmode M(L^-) \else $M(L^-)$\fi}}
\newcommand{\mlz}{{\ifmmode M(L^0) \else $M(L^0)$\fi}}
\newcommand{\lz}{{\ifmmode L^0 \else $L^0$\fi}}
\newcommand{\ev}{{\ifmmode GeV/c^2 \else $GeV/c^2$\fi}}
\newcommand{\tri}{{\ifmmode \triangleup \else $\triangleup$\fi}}
\newcommand{\unl}{{\ifmmode U_{lL^0} \else $U_{lL^0}$\fi}}\newcommand{\gL}{{\ifmmode g_L \else $g_{L}$\fi}}
\newcommand{\gR}{{\ifmmode g_R  \else $g_{R}$\fi}}
\newcommand{\gumu}{{\ifmmode \gamma^{\mu} \else $\gamma^{\mu}$\fi}}
\newcommand{\gunu}{{\ifmmode \gamma^{\nu} \else $\gamma^{\nu}$\fi}}
\newcommand{\gdmu}{{\ifmmode \gamma_{\mu} \else $\gamma_{\mu}$\fi}}
\newcommand{\gdnu}{{\ifmmode \gamma_{\nu} \else $\gamma_{\nu}$\fi}}
\newcommand{\stw}{{\ifmmode\sin^2\theta_W \else $\sin^{2}\theta_{W}$ \fi}}
\newcommand{\sws}{{\ifmmode \;\sin^2\theta_W  \else $\;\sin^{2}\theta_{W}$ \fi}}
\newcommand{\cws}{{\ifmmode \;\cos^2\theta_W  \else $\;\cos^{2}\theta_{W}$ \fi}}
\newcommand{\sw}{{\ifmmode \;\sin\theta_W  \else $\sin\theta_{W}$ \fi}}
\newcommand{\cw}{{\ifmmode \;\cos\theta_W  \else $\;\cos\theta_{W}$ \fi}}
\newcommand{\tw}{{\ifmmode \;\tan\theta_W  \else $\;\tan\theta_{W}$ \fi}}
\newcommand{\qq}{{\ifmmode q\overline{q} \else $q\overline{q}$\fi}}
\newcommand{\lR}{{\ifmmode l_R \else $l_R$\fi}}
\newcommand{\lL}{{\ifmmode l_L \else $l_L$\fi}}
\newcommand{\nt}{{\ifmmode \nu_{\tau} \else $\nu_{\tau}$\fi}}
\newcommand{\nuR}{{\ifmmode \nu_R  \else $\nu_R$\fi}}
\newcommand{\nuL}{{\ifmmode \nu_L  \else $\nu_L$\fi}}
\newcommand{\qR}{{\ifmmode g_R  \else $q_R$\fi}}
\newcommand{\qL}{{\ifmmode q_L  \else $q_L$\fi}}
\newcommand{\qRp}{{\ifmmode q_R'  \else $q_{R}$'\fi}}
\newcommand{\qLp}{{\ifmmode q_L'  \else $q_{L}$'\fi}}
\newcommand{\est}{{\ifmmode e^{\bf \ast} \else $e^{\bf \ast}$\fi}}
\newcommand{\lst}{{\ifmmode l^{\bf \ast} \else $l^{\bf \ast}$\fi}}
\newcommand{\must}{{\ifmmode \mu^{\bf \ast} \else $\mu^{\bf \ast}$\fi}}
\newcommand{\taust}{{\ifmmode \tau^{\bf \ast} \else $\tau^{\bf \ast}$ \fi}}
\newcommand{\pperp}{{\ifmmode p_t  \else $p_t$\fi}}
\newcommand{\et}{{\ifmmode E_t  \else $E_t$\fi}}
\newcommand{\xt}{{\ifmmode x_t  \else $x_t$\fi}}
\newcommand{\smumu}{{\ifmmode \sigma_{\mu\mu}  \else $\sigma_{\mu\mu}$ \fi}}
\newcommand{\eg}{{\ifmmode e\gamma  \else $e\gamma$\fi}}
\newcommand{\epem}{{\ifmmode e^+e^-  \else $e^+e^-$\fi}}
\newcommand{\lplm}{{\ifmmode L^+L^-  \else $L^+L^-$\fi}}
\newcommand{\pp}{{\ifmmode p\overline p  \else $p\overline p$\fi}}
\newcommand{\llz}{{\ifmmode L^0\overline{L}^0 \else $L^0\overline{L}^0$\fi}}
\newcommand{\epemt}{{\ifmmode e^+e^- \to  \else $e^+e^- \to$\fi}}
\newcommand{\eb}{{\ifmmode E_{beam}  \else $E_{beam}$\fi}}
\newcommand{\ip}{{\ifmmode pb^{-1}  \else $pb^{-1}$\fi}}
\newcommand{\upm}{{\ifmmode ^{\pm}  \else $^{\pm}$\fi}}
\newcommand{\de}{{\ifmmode ^{\circ}  \else $^{\circ}$ \fi}}
\newcommand{\appr}{{\ifmmode \sim \else $\sim$ \fi}}
\newcommand{\corresp}{{\ifmmode \stackrel{\wedge}{=} \else $\stackrel{\wedge}{=}$ \fi}}
\newcommand{\sqrts}{{\ifmmode \sqrt{s} \else $\sqrt{s}$\fi}}
\newcommand{\zz}{{\ifmmode Z^0  \else $Z^0$\fi}}
\newcommand{\mz}{{\ifmmode M_{Z}  \else $M_{Z}$\fi}}
\newcommand{\mzs}{{\ifmmode M_{Z}^2  \else $M_{Z}^2$\fi}}
\newcommand{\mw}{{\ifmmode M_{W}  \else $M_{W}$\fi}}
\newcommand{\mws}{{\ifmmode M_{W}^2  \else $M_{W}^2$\fi}}
\newcommand{\mh}{{\ifmmode M_{Higgs}  \else $M_{Higgs}$\fi}}
\newcommand{\gt}{{\ifmmode \Gamma_{tot} \else $\Gamma_{tot}$\fi}}
\newcommand{\msusy}{{\ifmmode M_{SUSY}  \else $M_{SUSY}$\fi}}
\newcommand{\msusys}{{\ifmmode M_{SUSY}^2  \else $M_{SUSY}^2$\fi}}
\newcommand{\su}{{\ifmmode SU(3)_C\otimes\- SU(2)_L\otimes\- U(1)_Y \else $SU(3)_C\otimes SU(2)_L\otimes U(1)_Y$\fi}}
\newcommand{\suthree}{{\ifmmode SU(3)_C  \else $SU(3)_C$\fi}}
\newcommand{\sutwo}{{\ifmmode  SU(2)_L\otimes U(1)_Y \else $SU(2)_L\otimes U(1)_Y$\fi}}
\newcommand{\taup} {{\ifmmode \tau_{proton} \else $\tau_{proton}$\fi}}
\newcommand{\as}{{\ifmmode \alpha_{s}  \else $\alpha_{s}$\fi}}
\newcommand{\mgut}{{\ifmmode M_{GUT}  \else $M_{GUT}$\fi}}
\newcommand{\mguts}{{\ifmmode M_{GUT}^2  \else $M_{GUT}^2$\fi}}
\newcommand{\mze} {{\ifmmode m_0        \else $m_0$\fi}}
\newcommand{\mha}{{\ifmmode m_{1/2}    \else $m_{1/2}$\fi}}
\newcommand{\mb} {{\ifmmode m_{b}    \else $m_{b}$\fi}}
\newcommand{\mt} {{\ifmmode m_{t}    \else $m_{t}$\fi}}
\newcommand{\mts} {{\ifmmode m_{t}^2    \else $m_{t}^2$\fi}}

\newcommand{\mtau}{{\ifmmode m_{\tau}  \else $m_{\tau}$\fi}}
\newcommand{\dpp}{{\ifmmode \delta_{pert} \else $\delta_{pert}$\fi}}
\newcommand{\dnp}{{\ifmmode\delta_{non-pert}\else$\delta_{non-pert}$\fi}}
\newcommand{\dew}{{\ifmmode \delta_{\rm EW}\else $\delta_{\rm EW}$\fi}}
\newcommand{\rt}{{\ifmmode R_{\tau}  \else $R_{\tau} $\fi}}
\newcommand{\rz}{{\ifmmode R_{Z}  \else $R_{Z} $\fi}}

\newcommand{\swb}{{\ifmmode \sin^2\theta_{\overline{MS}} \else $\sin^2\theta_{\overline{MS}}$\fi}}
\newcommand{\cwb}{{\ifmmode \cos^2\theta_{\overline{MS}} \else $\cos^2\theta_{\overline{MS}}$\fi}}

\newcommand{\AmS}{{\protect\the\textfont2 A\kern-.1667em\lower.5ex\hbox{M}\kern-.125emS}}

\newcommand{\mzero}{\rm m_0}
\newcommand{\mhalf}{\rm m_{1/2}}

\begin{document}

\begin{frontmatter}


\title{Excess of EGRET Galactic Gamma Ray Data interpreted as Dark Matter Annihilation}

\author{W. de Boer, M. Herold, C. Sander, V. Zhukov\corauthref{cor2}}
\corauth[cor2]{On leave from INPMSU, Moscow.}
\address{Institut f\"ur Experimentelle Kernphysik\\
Universit\"at Karlsruhe (TH), P.O. Box 6980, 76128 Karlsruhe, Germany\\}
\author{A.V. Gladyshev, D.I.  Kazakov}
\address{Bogoliubov Laboratory of Theoretical Physics, Joint Institue for Nuclear Research\\
141 980 Dubna, Moscow Region, Russia}
\begin{abstract}
The diffuse galactic EGRET gamma ray data
 show a clear excess for energies above 1 GeV in comparison
with the expectations from conventional galactic models. The excess is seen with the same
spectrum in all sky directions, as expected for Dark Matter (DM) annihilation. This
hypothesis is investigated in detail. The energy spectrum of the excess is used to limit
the WIMP mass to the 50-100 GeV range, while the sky maps are used to determine the halo
structure, which is consistent with a triaxial isothermal halo with additional enhancement
of Dark Matter in the disc. The latter is strongly correlated with the ring of stars around
our galaxy at a distance of 14 kpc, thought to originate from the tidal disruption of a
dwarf galaxy. It is shown that this ring of DM with a mass of $\approx 8\cdot
10^{10}~M_\odot$ causes the mysterious change of slope in the rotation curve
  at $R=1.1R_0$ and the large local surface density of the disc. The total mass of the halo
  is determined to be $3\cdot 10^{12}~M_\odot$.
  A cuspy  profile is definitely excluded to describe the gamma ray data.
These signals of Dark Matter Annihilation are compatible with Supersymmetry for boost
factors of 20 upwards and have a statistical significance of more than $10\sigma$ in
comparison with the conventional galactic model. The latter combined with  all  features
mentioned above provides an intriguing hint that the EGRET excess is indeed  a signal from
Dark Matter Annihilation.

\end{abstract}

\end{frontmatter}

\section{Introduction}
Cold Dark Matter (CDM) makes up 23\% of the energy of the universe, as deduced from the
WMAP measurements of the temperature anisotropies in the Cosmic Microwave Background, in
combination with data on the Hubble expansion and the density fluctuations in the
universe~\cite{wmap}. The nature of the CDM is unknown, but one of the most popular
explanation for it is the neutralino, a stable neutral particle predicted by
Supersymmetry~\cite{lspdm,jungman}. The neutralinos are spin 1/2 Majorana particles, which
can annihilate into pairs of Standard Model (SM) particles. The stable decay and
fragmentation products are neutrinos, photons, protons, antiprotons, electrons and
positrons. From these, the protons and electrons disappear in the sea of many matter
particles in the universe, but the photons and antimatter particles may be detectable above
the background, generated by  particle interactions. Such  searches for indirect Dark
Matter detection have been actively pursued, see e.g the review by Bergstr$\rm
\ddot{o}$m\cite{bergstrom} or more recently by Bertone, Hooper and Silk
\cite{Bertone:2004pz}. In a previous paper we studied the Dark Matter Annihilation (DMA)
yield of positrons, antiprotons and gamma rays from the center of the Galaxy  using the
cuspy NFW halo profile expected from N-body simulations\cite{us}. It was found that the DMA
contribution to all three channels could be well described by the data for an annihilation
cross section determined from the WMAP measurement of the relic density and using a simple
propagation model. In this paper we extend this analysis to include gamma rays from {\it
all} sky directions. Gamma rays have the advantage that they point back to the source and
do not suffer energy losses, so they are the ideal candidates to trace the dark matter
density, if one assumes the  boost factor, representing  local density fluctuations of the
DM, to be similar in all directions. The charged components interact with galactic matter
and scatter on magnetic turbulences, so they do not point back to the source.   They can
only be studied by including the DM as a source function in the standard galactic
propagation model as implemented in the GALPROP code\cite{galprop}. This has been done, but
its discussion is outside the scope of the present paper.
 The present analysis on diffuse galactic gamma rays differs from previous ones (see e.g. \cite{previous}) by
considering simultaneously the complete sky map {\it and} the energy spectrum, which allows
us to constrain both the halo distribution {\it and} the WIMP mass.
The constraint on the WIMP annihilation cross section from WMAP is discussed in Section 2,
while the constraints on the mass and the DM
halo profile from the EGRET excess are discussed in Sections 3 and 4. The compatibility of the results with Supersymmetry is
discussed in Section 5 and the summary is given in
Section 6.

\section{Annihilation Cross section Constraints from WMAP}
In the early universe all particles were produced abundantly and
were in thermal equilibrium through annihilation and production
processes. The time evolution of the number density of the
particles is given by the Boltzmann equation, which can be written
for WIMPS, denoted by $\chi$, as: \bq \frac{dn_\chi}{dt}+3Hn_\chi=-<\sigma
v>(n_\chi^2-n_\chi^{eq 2}), \eq where H is the Hubble expansion
rate, $n_\chi$ is the actual number density, $n_\chi^{eq}$ is the
thermal equilibrium number density, $<\sigma
v>$ is thermally averaged value of the total annihilation cross
section times the relative velocity of the annihilating
neutralinos. The Hubble term takes care of the decrease in number
density because of the expansion, while the first term on the
right hand side represents the decrease due to annihilation and
the second term represents the increase through creation by the
inverse reactions.

At temperatures below the mass of the WIMP's the number
density drops exponentially. The annihilation rate $\Gamma=<\sigma
v> n_\chi$ drops exponentially as well, and if it drops below the
expansion rate, the WIMP's cease to annihilate. They fall out
of equilibrium (freeze-out) at a temperature of about $m_\chi/22$
~\cite{kolb} and a relic cosmic abundance remains.

For the case that $<\sigma v>$ is energy independent, which is a
good approximation in case there is no coannihilation,
the present mass density in units of the
critical density is given by~\cite{jungman}: \bq \Omega_\chi
h^2=\frac{m_\chi n_\chi}{\rho_c}\approx (\frac{2\cdot 10^{-27}
cm^3 s^{-1}}{<\sigma v>})\label{wmap}.\eq One observes that the
present relic density is inversely proportional to the
annihilation cross section at the time of freeze out, a result
independent of the WIMP   mass (except for logarithmic
corrections). For the present value of $\Omega_\chi h^2=0.113\pm0.009$ the
thermally averaged total cross section at the freeze-out
temperature of $m_\chi/22$ must have been around $2\cdot 10^{-26} {\rm cm^3s^{-1}}$.
All possible enhancements (boost factors) of the annihilation rate will be calculated with respect
to this generic cross section, which basically only depends on the value
of the Hubble constant in absence of resonances and if coannihilation with
other particles can be neglected.

\begin{table}
\begin{center}
 \begin{tabular}{cccc} \hline
 Region & Longitude $l$ & Latitude $|b|$ & Description\\\hline
 A & 330-30 & 0-5 & Inner Galaxy\\
 B & 30-330 & 0-5 & Galactic disc without inner Galaxy\\
 C & 90-270 & 0-10 & Outer Galaxy\\
 D & 0-360 & 10-20 & low longitude \\
 E & 0-360 & 20-60 & high longitude\\
 F & 0-360 & 60-90 & Galactic Poles\\
 \hline
 \end{tabular}
 \end{center}
 \caption[skyregions]{The six sky regions used in this analysis.}
 \label{t1}
 \end{table}
\begin{figure}[t]
\begin{center}
 \includegraphics [width=0.44\textwidth,clip]{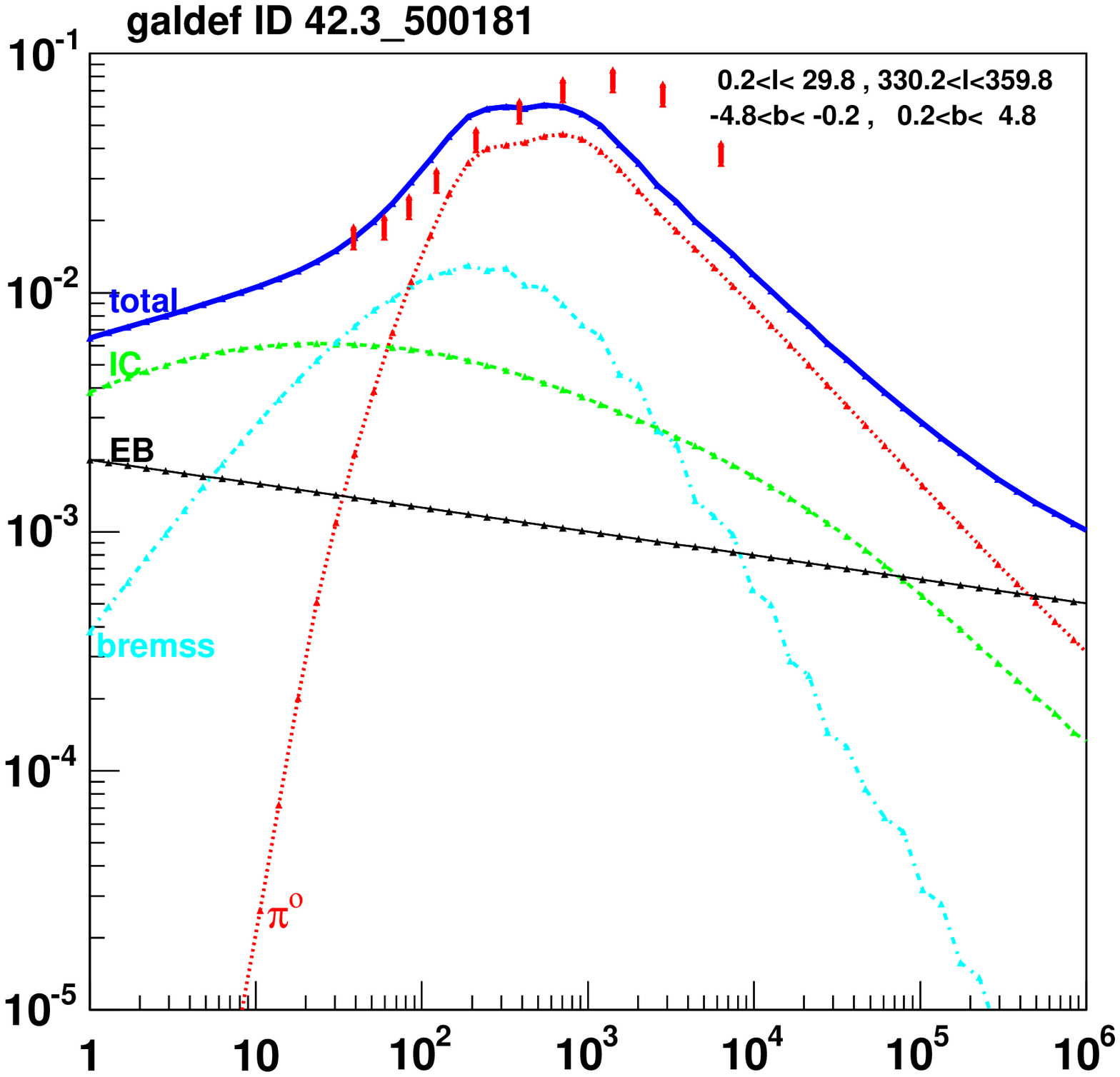}
 \includegraphics [width=0.4\textwidth,clip]{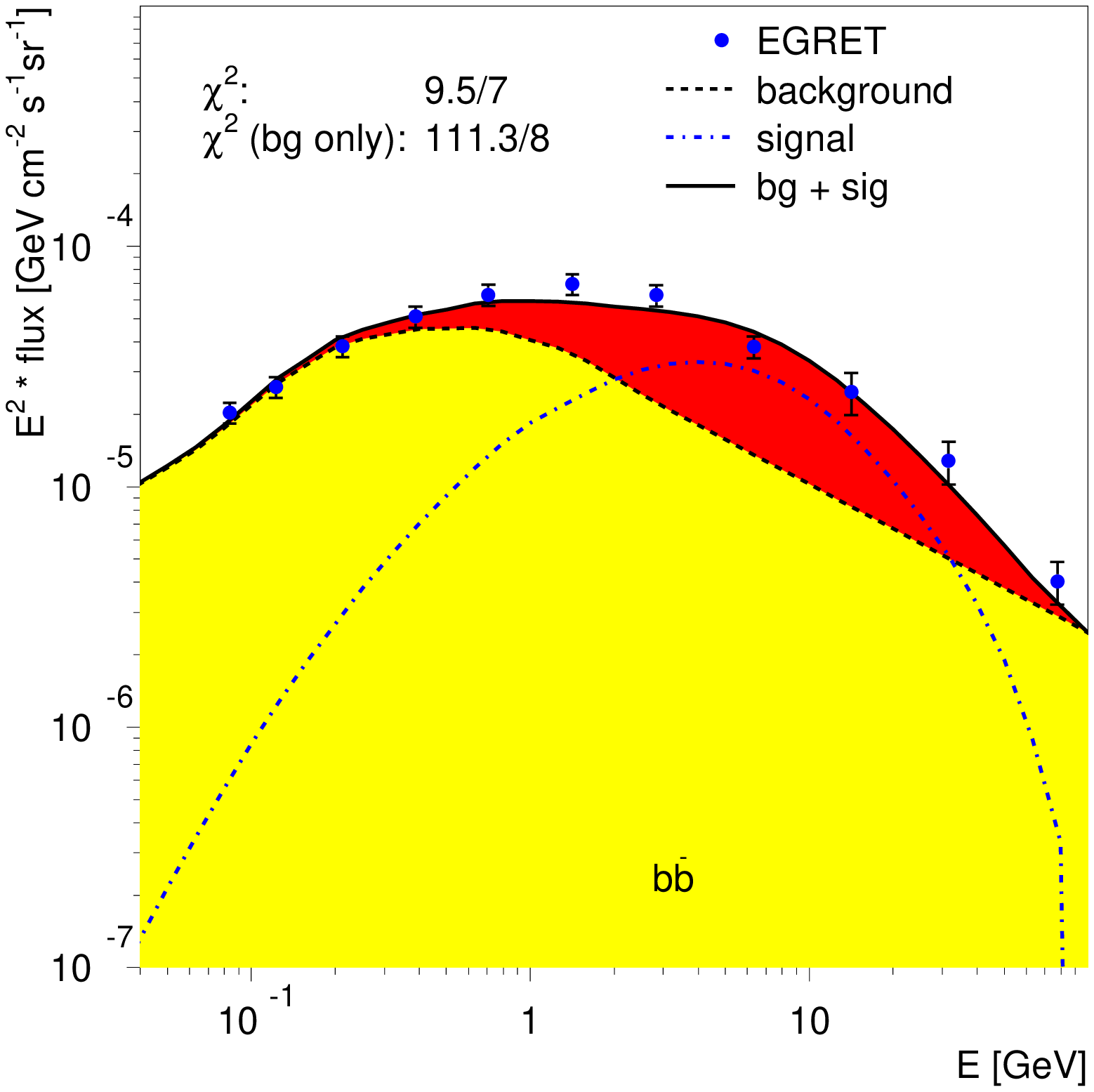}
 \caption[]{Left:
 The diffuse gamma-ray energy spectrum of the inner  Galaxy as calculated
 with the conventional  GALPROP model in comparison with EGRET data.
Right: as on the left, but with an additional component from Dark Matter Annihilation.
The EGRET data above 10 GeV are from Ref. \cite{strong_egret}}
 \label{gamma_A}
\end{center}
\end{figure}
\begin{figure}[t]
\begin{center}
 \includegraphics [width=0.45\textwidth,clip]{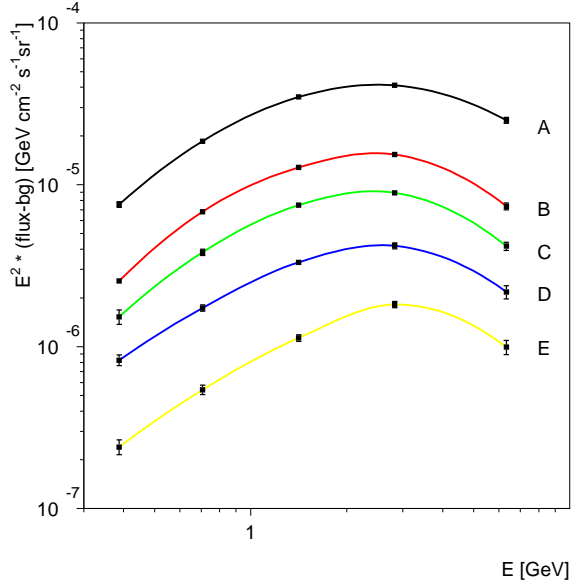}
 \caption[]{
 The excess of the EGRET data for the various regions of Table \ref{t1}, determined
 simply by plotting the differences between the conventional GALPROP model and the EGRET data.
 One observes the same spectral shape for all regions, indicating a common source
 for the excess.
 \label{excess}}
\end{center}
\end{figure}

\section{Global Fits to the spectral shape of the diffuse gamma rays}\label{spectrum}
 Galactic gamma ray have been extensively studied during the nine years of data taking by the EGRET
satellite on the Compton Gamma Ray Observatory. These data show a clear excess of diffuse
gamma rays above the background from conventional galactic sources for energies above 1 GeV\cite{strong_egret},
 as shown on the left hand side of Fig. \ref{gamma_A}.

 Of course there are also many sources of diffuse gamma rays in the galaxy, so disentangling the
annihilation signal is at first glance not easy.
The main sources of conventional background are: a)
decays from $\pi_0$ mesons produced in nuclear interactions (mainly inelastic proton-proton
or p-He collisions); b) inverse Compton scattering of electrons on photons (e.g. from star
light or cosmic microwave background); c) Bremsstrahlung from electrons. WIMP's are expected
to annihilate in fermion-antifermion pairs, so a large fraction will annihilate into quark
pairs, which produce typically 30-40 photons per annihilation in the fragmentation process
(mainly from $\pi_0$ decays as in nuclear interactions). However, the photons from DMA  are
expected to have a significantly different spectrum than the ones from nuclear
interactions. This can be easily seen as follows: the WIMP's are strongly non-relativistic,
so they annihilate almost at rest. Therefore the quarks from DMA are almost mono-energetic
with  an energy approximately equal to the WIMP mass. This results in a rather energetic
spectrum with a sharp cut-off of the photons at twice the WIMP mass. Such a spectrum
deviates considerably from the photons from nuclear interactions, which peaks at 70 MeV and
falls off according to a power law.  This difference allows one to fit simultaneously only
the {\it shapes} of the spectra from background and DMA, thus treating the absolute
normalizations as free parameters, which is an advantage, since
the shapes of the spectra are much better known
than the absolute fluxes.

The left hand side of
Fig. \ref{gamma_A} shows the diffuse gamma ray data in region A together with the data from
the background processes. The sum of these contributions is shown by the solid line and
clearly fails to describe the EGRET data, shown by the points with vertical error bars.
Providing a better fit to the data without assuming DMA requires a harder proton spectrum.
However, this is constrained by the number of antiprotons. Leaving both the electron
spectrum and proton spectrum free, i.e. assuming that the averaged spectra in our Galaxy
can be different from the locally measured spectra, allows a reasonable description of all
data\cite{strong_egret}. In this so-called optimized model the locally measured proton
(electron) spectrum is about a factor 2(4) below the averaged galactic one.
As an alternative, we have been investigating if Dark Matter annihilation can be
responsible for the excess. The fit to  the EGRET data including DMA is shown on the right
hand side of Fig. \ref{gamma_A} for a WIMP mass of 90 GeV.
The shape of the background is taken from the GALPROP program,
which provides a detailed simulation of our galaxy\cite{galprop}.
The shape of the DMA curve was
taken from the DarkSusy program\cite{darksusy}, which uses the Pythia program\cite{pythia} for the
fragmentation of the quarks. Although DarkSusy uses the supersymmetric neutralinos as
WIMP's, the shape of the curve is generic for any WIMP annihilating at rest into quark
pairs, which fragment according to the LUND string model\cite{pythia}. This model has been
well tested in many reactions and the small differences between different quark flavours
can be compensated by a different WIMP mass. The EGRET excess up to 120 GeV is
well described by the contribution from DMA. Here  data for energies between 10 and
120 GeV (last three data points) were included\cite{strong_egret}. The EGRET data
have only been calibrated up to 10 GeV
in a test beam and the extrapolation to higher energies would require a  detailed Monte
Carlo simulation for the backsplash between calorimeter and veto counters, which has not yet been done.
 The more reliable data below 10 GeV
yields a slightly lower WIMP mass of 50  GeV. The range 50 to 90 GeV is acceptable for all
fits and can be used to estimate the uncertainty in the WIMP mass, for which 70 GeV is
used. It should be noted that in Supersymmetry the  dominant annihilation mode is
into $b\overline{b}$ quarks pairs\cite{us}, which are known to have a hard fragmentation function.
Annihilation e.g. into $W^+W^-$ yields a softer gamma spectrum, which can be compensated by
a  heavier WIMP mass (about 30 GeV heavier), so the WIMP mass range  given above should be
considered a lower limit.

\begin{figure}
\begin{center}
 \includegraphics [width=0.35\textwidth,clip]{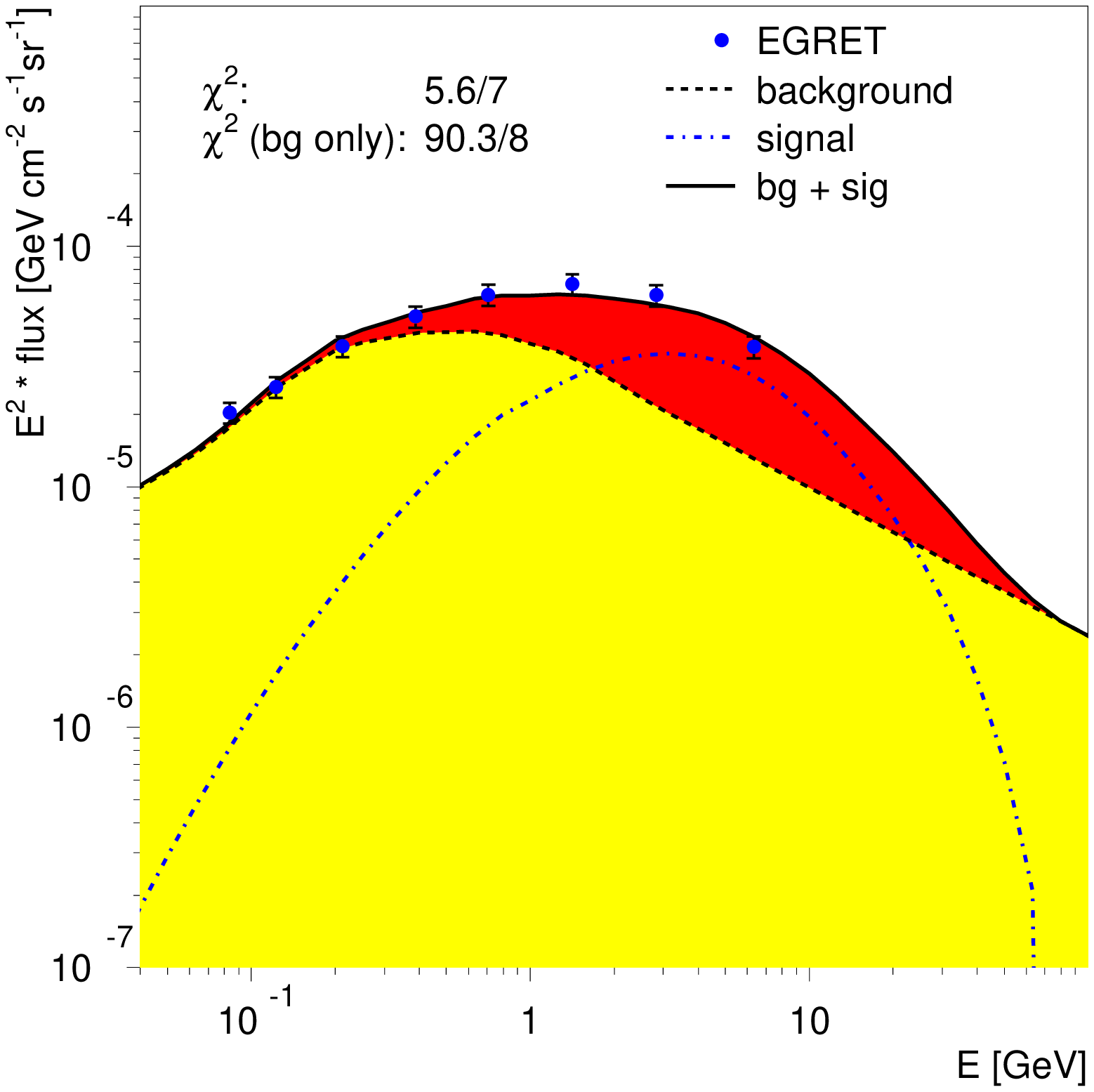}
 \includegraphics [width=0.35\textwidth,clip]{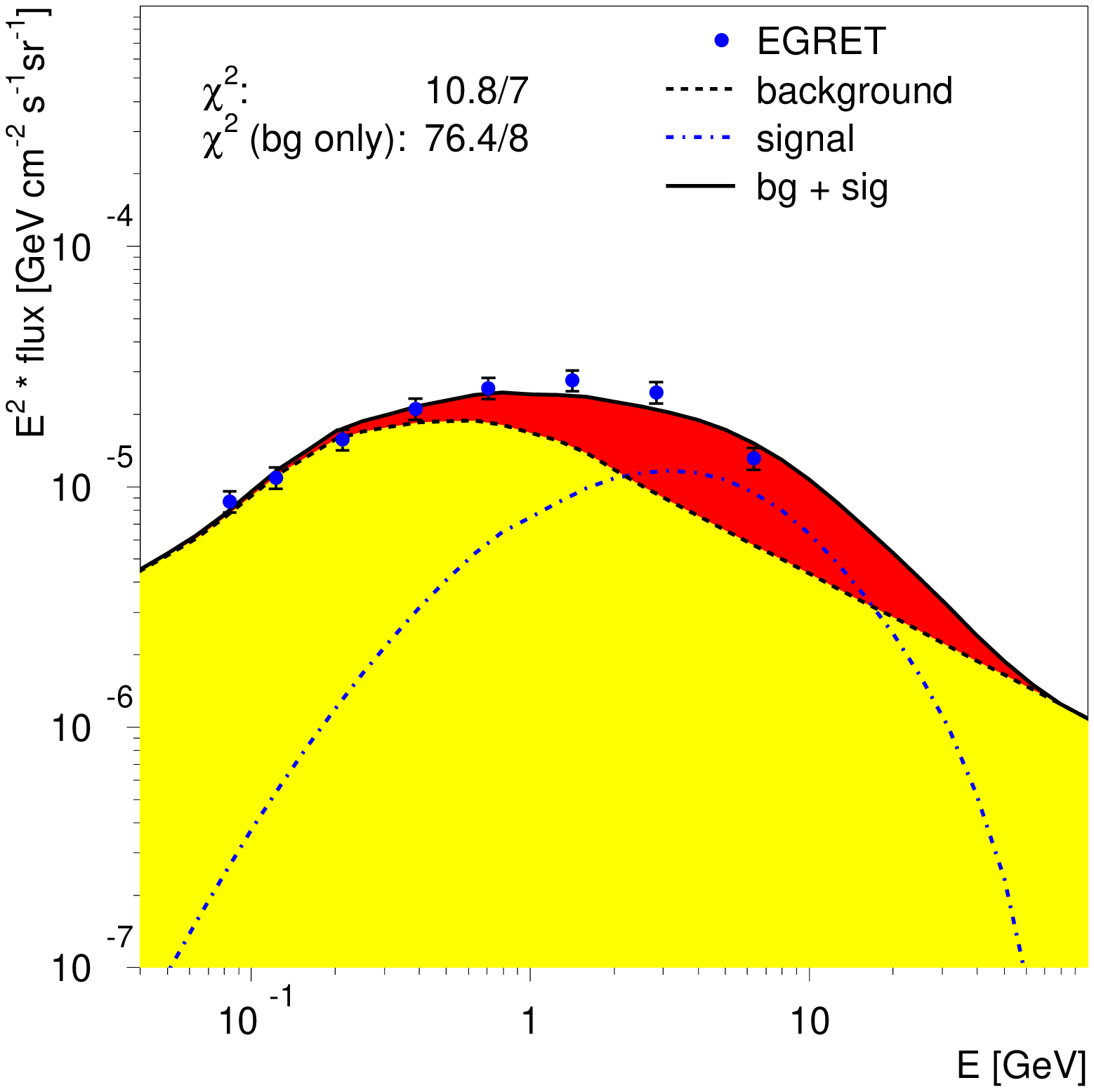}
 \includegraphics [width=0.35\textwidth,clip]{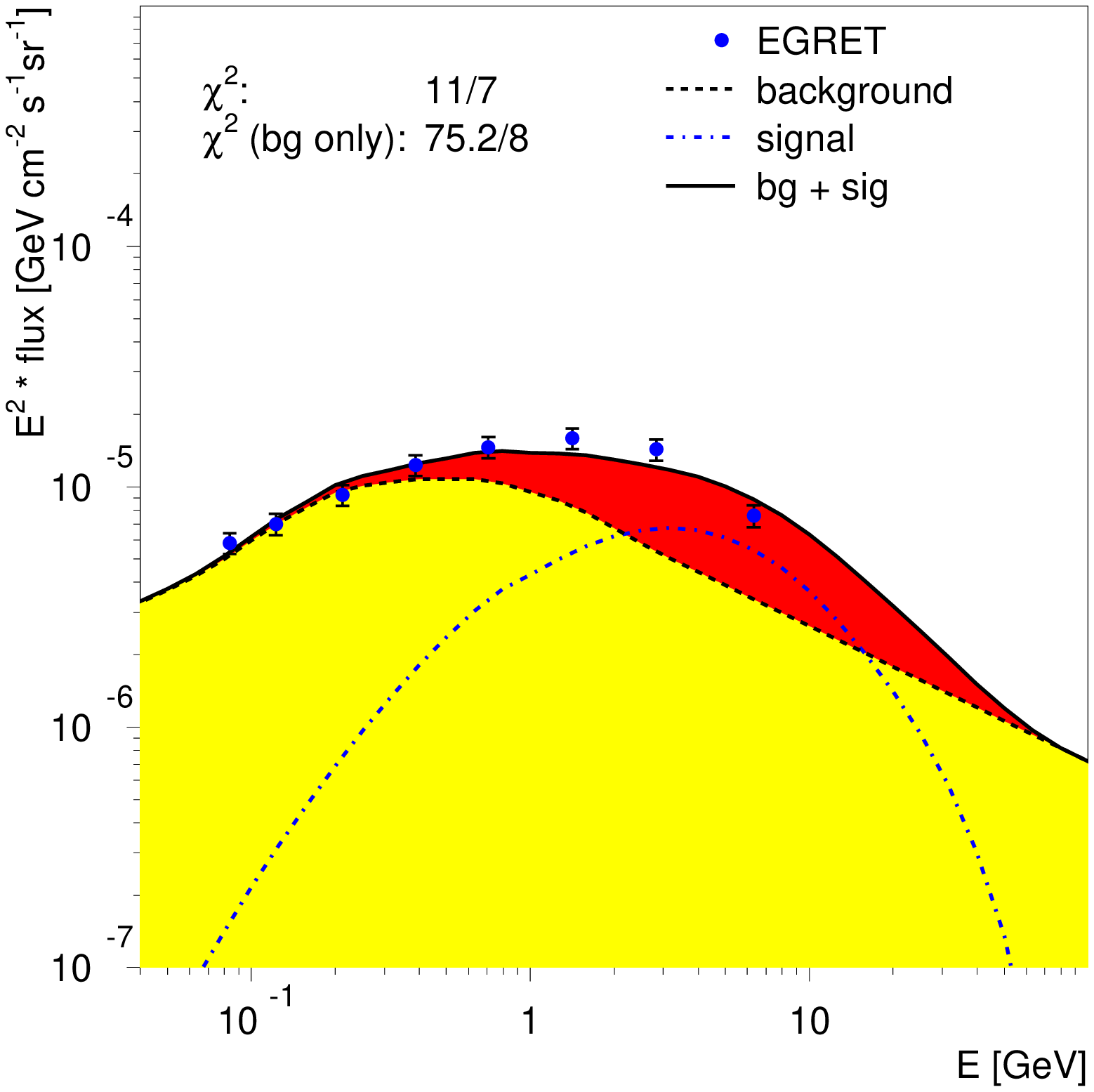}
 \includegraphics [width=0.35\textwidth,clip]{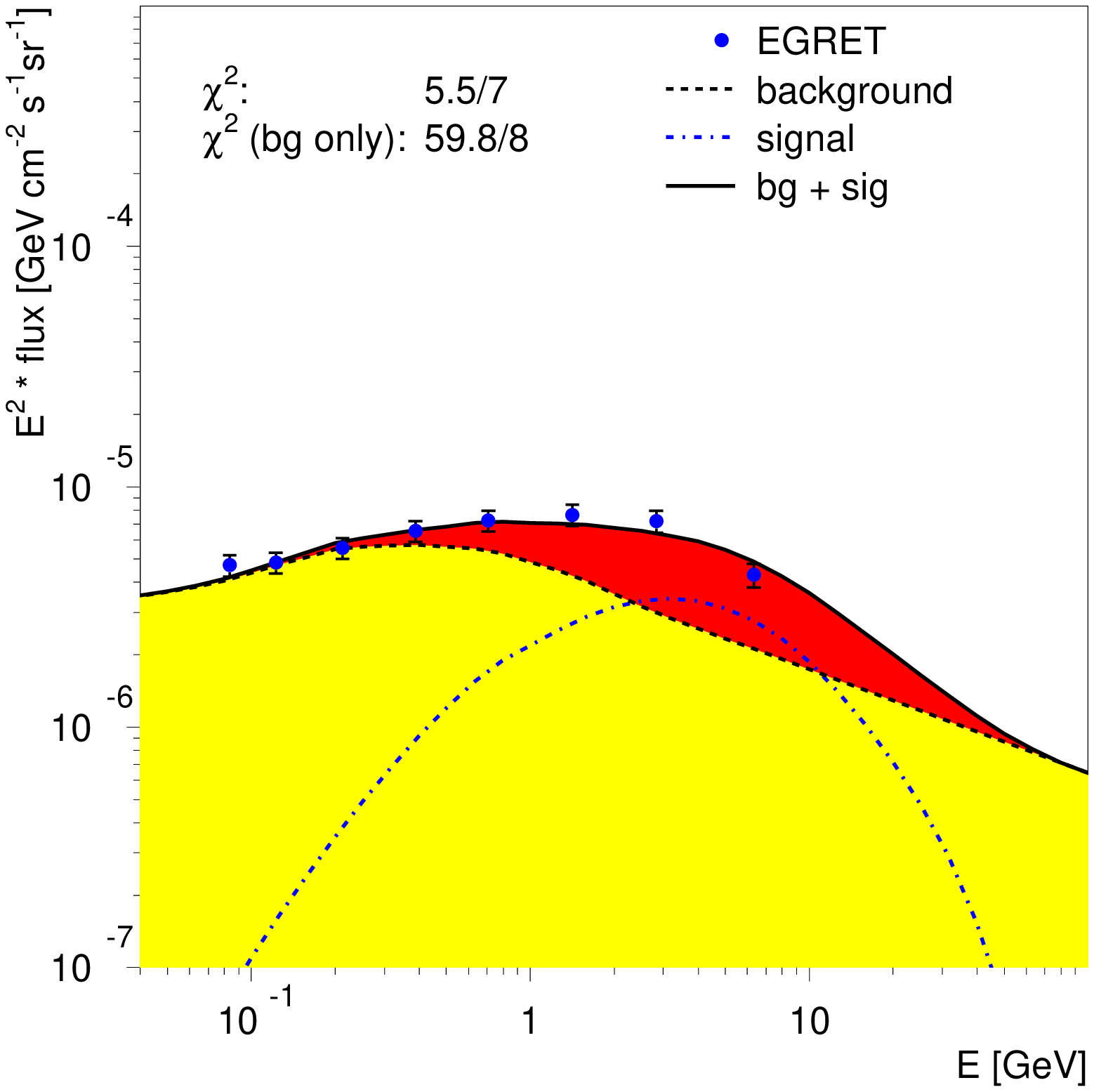}
 \includegraphics [width=0.35\textwidth,clip]{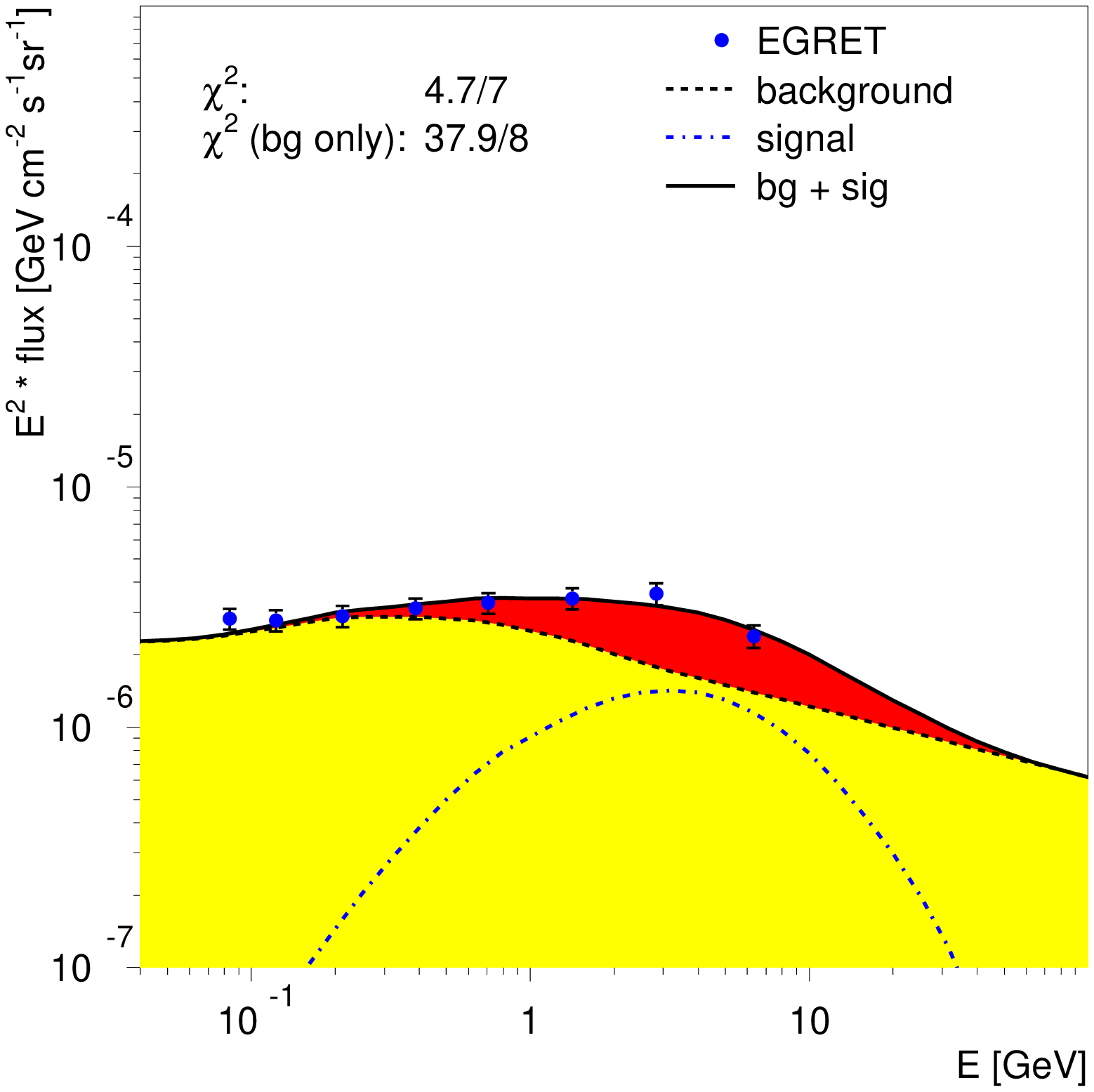}
 \includegraphics [width=0.35\textwidth,clip]{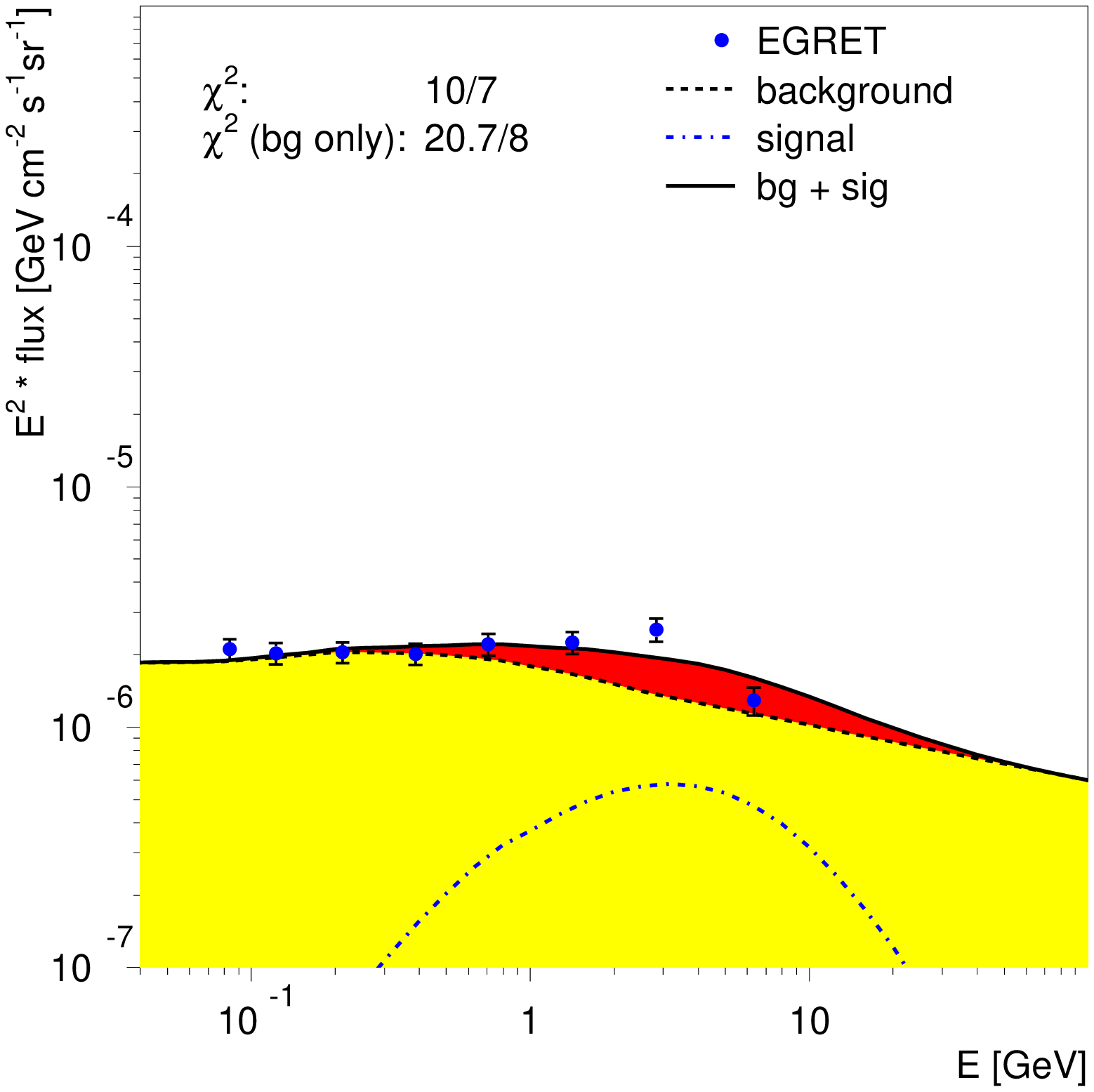}
 \caption[]{
 The diffuse gamma-ray energy spectrum of the 6 angular regions given in Table \ref{t1},
  as measured by the EGRET satellite.
 The contributions from the background and the neutralino annihilation signal have
  been indicated by the light (yellow) and dark (red) shaded area, respectively.
 The fitted curves do not rely on the absolute predictions, but only assume the
{\it shapes} to be known. The dotted line is the shape of the DM annihilation
for a WIMP mass of 70 GeV and assuming annihilation into $b\overline{b}$ to be dominant.
The normalizations of background and DM annihilation
 are obtained from the fit to the data.
 \label{gamma_us}}
\end{center}
\end{figure}

The excess is observed in all sky directions with the same spectral shape, as shown in Fig.
\ref{excess} for the sky regions tabulated in  Table \ref{t1}). The same spectrum in all
sky directions is the hall mark of DMA and practically excludes the possibility that the
excess originates from unknown point sources, especially since most point sources have a
much softer spectrum than the observed high energy excess. Including all the known point
sources in the analysis does not change the results in any significant way. E.g. the many
sources in the direction of the galactic center increase the gamma ray flux in that
direction by less than 20\%, which only changes the normalization factor somewhat, but the
shape of the spectrum is hardly changed for the large angular regions considered here. Only
the statistical errors have been plotted in Fig. \ref{excess}, since the common
normalization error would only shift the whole curve without changing  its shape. The
largest excess comes from the galactic centre (region A), but at large latitudes (regions D
and E) there exists still a strong signal, as expected from DMA. The contributions from
background and signal in the various regions are shown in Fig. \ref{gamma_us}. Excellent
fits with free normalizations of DMA signal and background are obtained for all regions.
The DMA signal contributes differently for the different sky directions, as expected from
the fact that the halo profile has a maximum towards the centre and the annihilation rate
is proportional to the DM density squared.
Various proposed shapes of halo profiles will be discussed in the next section followed by
a determination of the halo parameters.

\section{Halo Profiles}\label{halo}

A survey of the optical rotation curves of 400 galaxies shows that the halo distributions
of most of them can be fitted either with the Navarro-Frank-White (NFW) or the
pseudo-isothermal profile\cite{Jimenez:2002vy}. The halo profiles can be parametrized as follows:
 $$\rho (r)=\rho_0\cdot (\frac{r}{a})^{-\gamma}\left[1+
 (\frac{r}{a})^\alpha\right] ^{\frac{\gamma-\beta}{\alpha}},$$
 where $a$ is a scale radius and the slopes $\alpha$, $\beta$ and
 $\gamma$ can be thought of as the radial dependence at $r\approx a$, $r>>a$ and $r<<a$, respectively.
The spherical profile can be somewhat flattened in two directions to form a triaxial halo.
N-body simulations suggest that  the ratio of the short (intermediate) axis to the major
axis is typically above 0.5 (0.7)\cite{Dubinski:1991bm,Jing:2002np}.

The cuspy NFW profile\cite{nfw} is defined by  $(\alpha, \beta, \gamma)$ =(1,3,1) for a
scale $a=10$ kpc, while the Moore profile with $\gamma=1.5$ is even more cuspy\cite{moore}.
The isothermal profile with  $(\alpha, \beta, \gamma)$ =(2,2,0) has no cusp ($\gamma=0$),
but a core which is taken to be the size of the inner galaxy, i.e. $a=4$ kpc and  $\beta=2$
implies a flat rotation curve.  There are several issues concerning the distribution of
dark matter in the galaxies:
\begin{itemize}
\item The NFW profile shows a strong increase in density near the center of the universe
($\rho\propto 1/r^\gamma$). However, the density profile in the inner parts of dwarf
galaxies and the rotation curves of low surface brightness galaxies (LSB), which presumably
consists largely of dark matter, point to a flat density profile near the
center\cite{blok}. A summary of the present discussions can be found e.g. in Ref.
\cite{primack,Merrifield:2003hz,mazzei}. Resonant interactions between a rotating bar of
visible matter and the dark matter may modify the cusp in barred galaxies and even cause a
``halo'' bar\cite{athanassoula,weinberg}.
 \item
 It is not clear if the dark matter is homogeneously distributed or has a clumpy character.
In N-body simulations the structure of the universe unfolds in a hierarchical manner,
starting with the growing of the random initial density fluctuations in the cosmic
microwave background and building up larger structure through mergers of smaller clusters.
This can lead to many streams of dark matter and an enhancement in the annihilation rate as
compared to a homogeneous distribution, since DMA $\propto\rho^2$. This enhancement
(boostfactor)  can be determined from a fit to the data.

\item It is usually assumed that DM does not interact with the visible matter, so many
N-body simulation only consider the DM contribution. However, at the center of a spiral
galaxy the gravitational potential is completely dominated by the visible matter and the DM
halo will adjust to it. This adiabatic compression can lead to an enhancement of the DM
density by factors of a few near the center of the galaxy\cite{adiabatic}, which in turn
might be distributed to larger radii by the resonant interactions mentioned above. \item
The DM usually forms sheets and filaments by gravitational collapse. Galaxies are formed
along these topological structures in the hierarchical manner discussed before, which leads
to correlations in the orientation of galaxy clusters\cite{faltenbacher,kasun}, but   can lead to
anisotropic infall at the galactic scale as well, as shown by N-body simulations by Aubert
et al. \cite{aubert,steinmetz}. An anisotropic infall along a filamentary structure can increase the
density in the plane, preferentially at large radii,  up to 100\% but is on average more
 like 15\% \cite{aubert}.
 In our galaxy the observed ring of stars at a radius of about 15 kpc might be an example
 of such an anisotropic infall\cite{yanny,ibata,helmi}.
\end{itemize}
The mechanisms above all lead to an enhancement of the Dark Matter density in the plane of
the disc, but N-body simulations do not have enough resolution to predict the actual
distribution. Therefore the possible enhancement of DM density in the disc was parametrized
by a set of Gaussian shaped rings in the galactic plane in addition to the expected
triaxial profile for the DM halo. At least two rings should be envisaged: one ``outer''
ring for the enhancement from anisotropic infall at larger radii and one ``inner'' ring
from the adiabatic compression by the gravitational potential of the visible matter. The
actual parameters of the Gaussian rings can be determined from a fit to the data. The n-th
ring is parametrized by a radius $R_n$ with a Gaussian width $\sigma_{Rn}$ in radius and
$\sigma_{zn}$ in height above the plane. So the total halo profile can be written as:
\begin{equation}
\rho_{\chi} (\tilde r)  = \rho_0  \left( \frac{R_0}{\tilde r} \right)^{\gamma} \left[
\frac{1 + \left( \frac{\tilde r}{a} \right)^{\alpha}} {1 + \left( \frac{R_0}{a}
\right)^{\alpha}}  \right]^{\frac{\gamma - \beta}{\alpha}} + \sum_{n=1}^N \rho_n  \exp
\left(-\frac {\left( \tilde{r}_{gc}-{Rn}\right)^2}{2
\sigma_{R_n}^2}-\frac{\left(z_n\right)^2}{2 \sigma_{z_n}^2} \right) \label{halo1}
\end{equation}
with
\begin{equation}
\tilde{r} = \sqrt{\frac{x^2}{a^2}+\frac{y^2}{b^2}+\frac{z^2}{c^2}}, \hspace*{1cm}
\tilde{r}_{gc} = \sqrt{\frac{x^2}{\tilde{a}^2}+\frac{y^2}{\tilde{b}^2}},
\end{equation}
and $a>b>c$ ($\tilde{a}>\tilde{b}$)are the principal axis of the triaxial halo profile
(ring).  The maximal neutralino density of a ring $\rho_{n}$ is reached in the galactic
plane ($z = 0$) at a distance from the galactic center $\tilde{r}_{gc}=R_n$. At present
only two rings are fitted, i.e. N=2. The radius of the ring can be determined from the
longitude profile in the plane of the galaxy, i.e. at low latitudes, because the solar
system is not at the center, so if the density is constant along the ring, the different
longitudes yield different fluxes, which depend on the radius, orientation and ellipticity
of the ring. The extent of the ring above the plane can be obtained from the longitude
distribution  for higher latitudes.
\begin{figure}
\begin{center}
 \includegraphics [width=0.3\textwidth,clip]{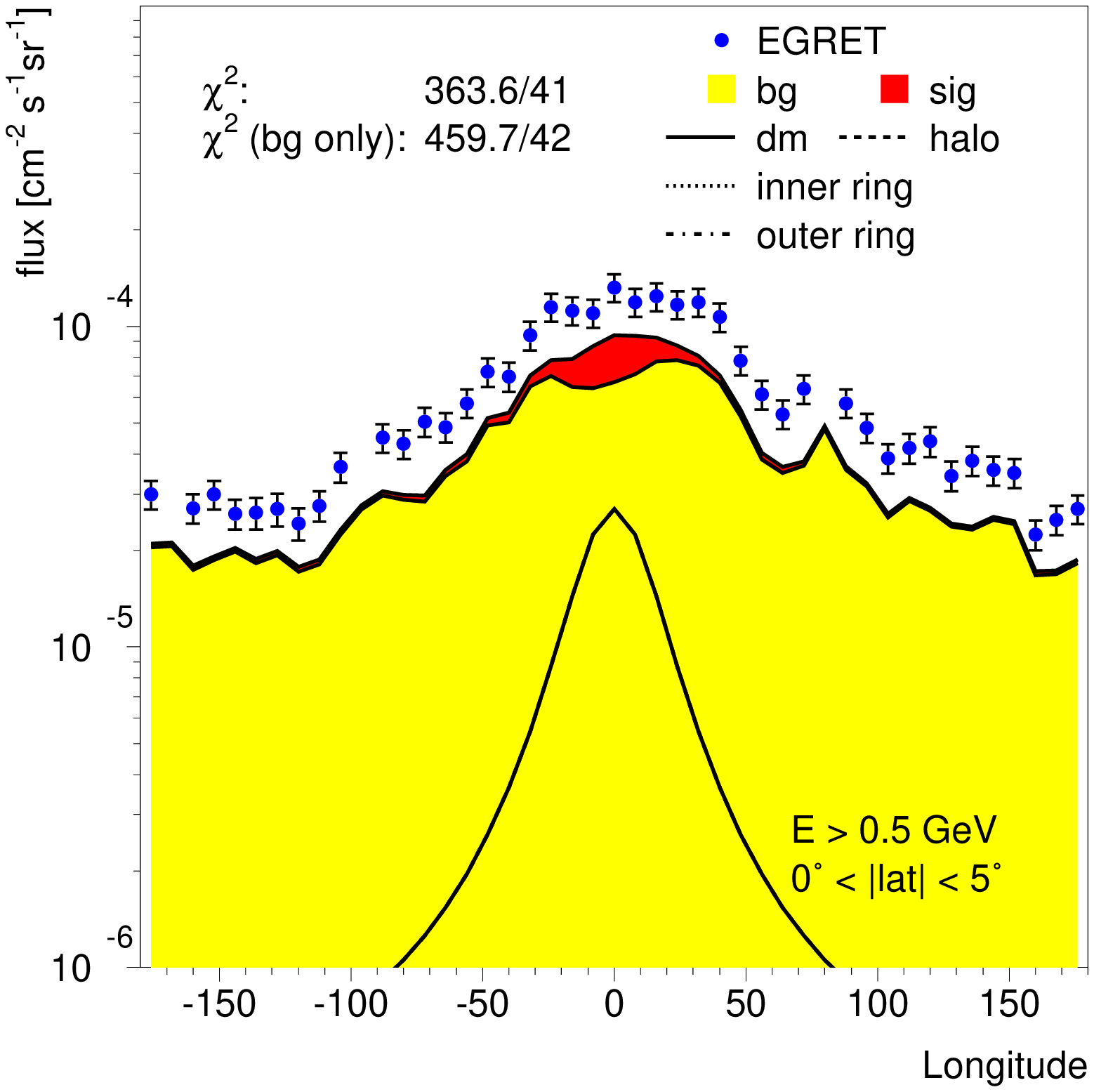}
 \includegraphics [width=0.3\textwidth,clip]{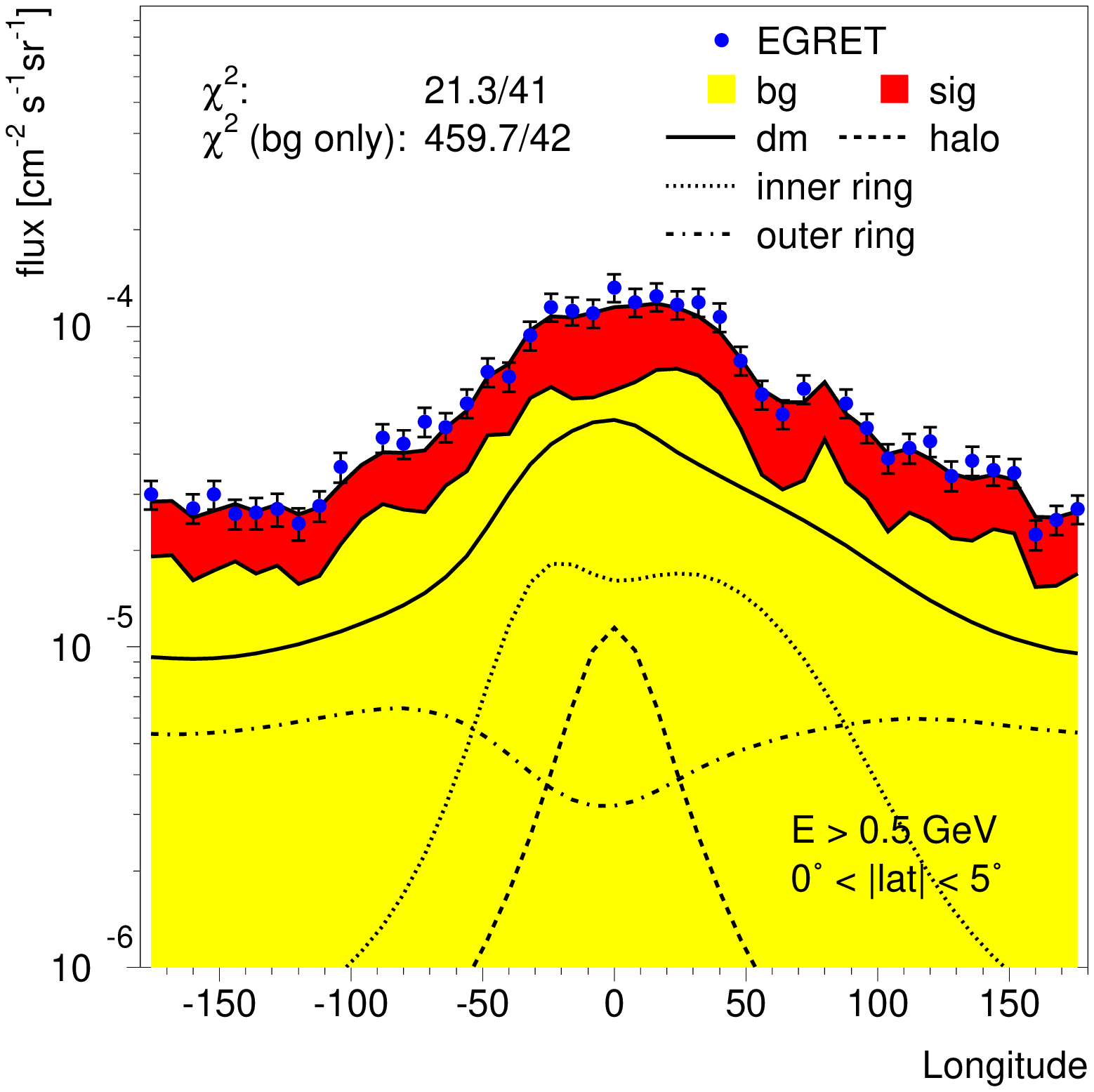}\\
  \includegraphics[width=0.3\textwidth,clip]{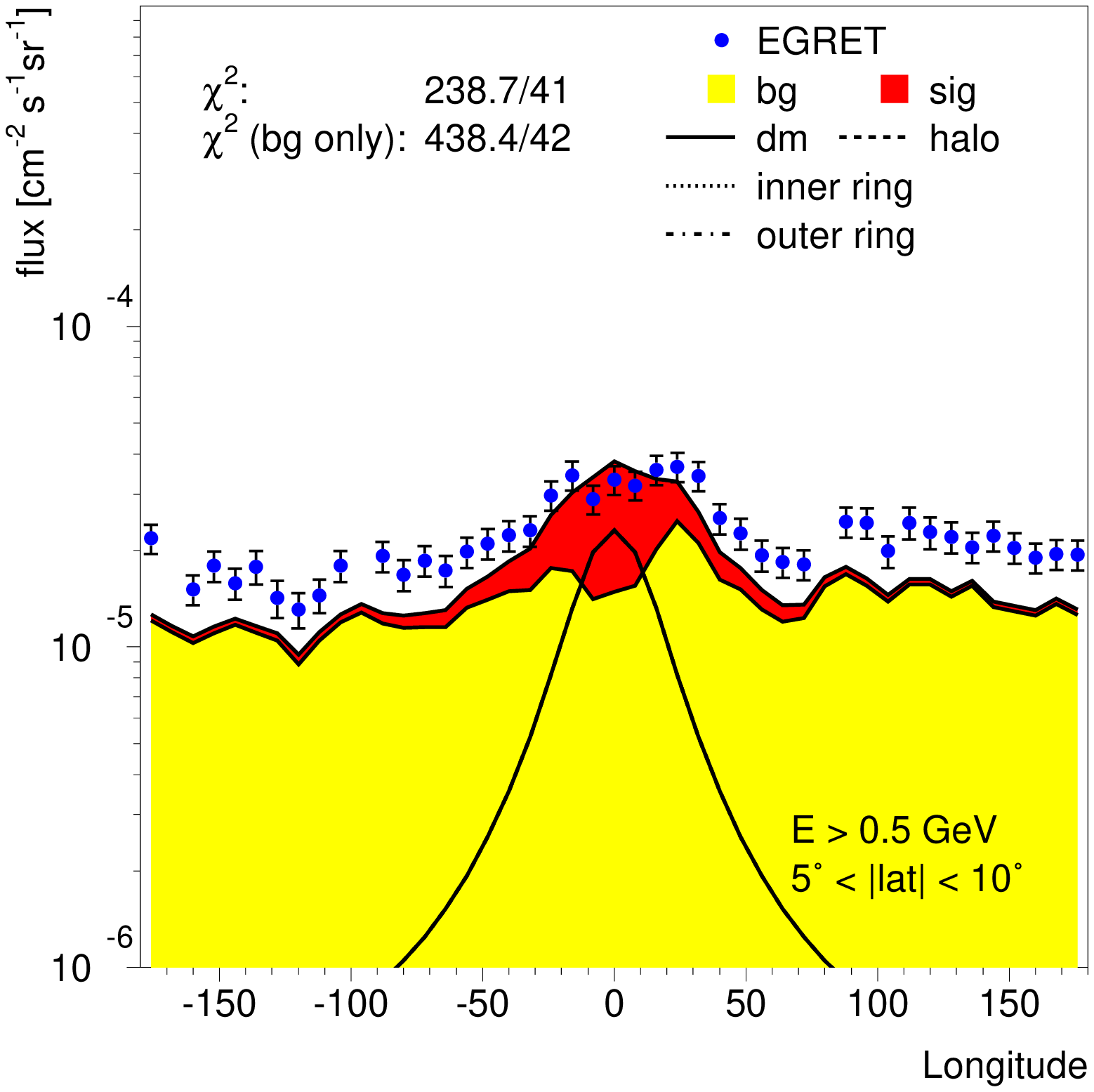}
 \includegraphics [width=0.3\textwidth,clip]{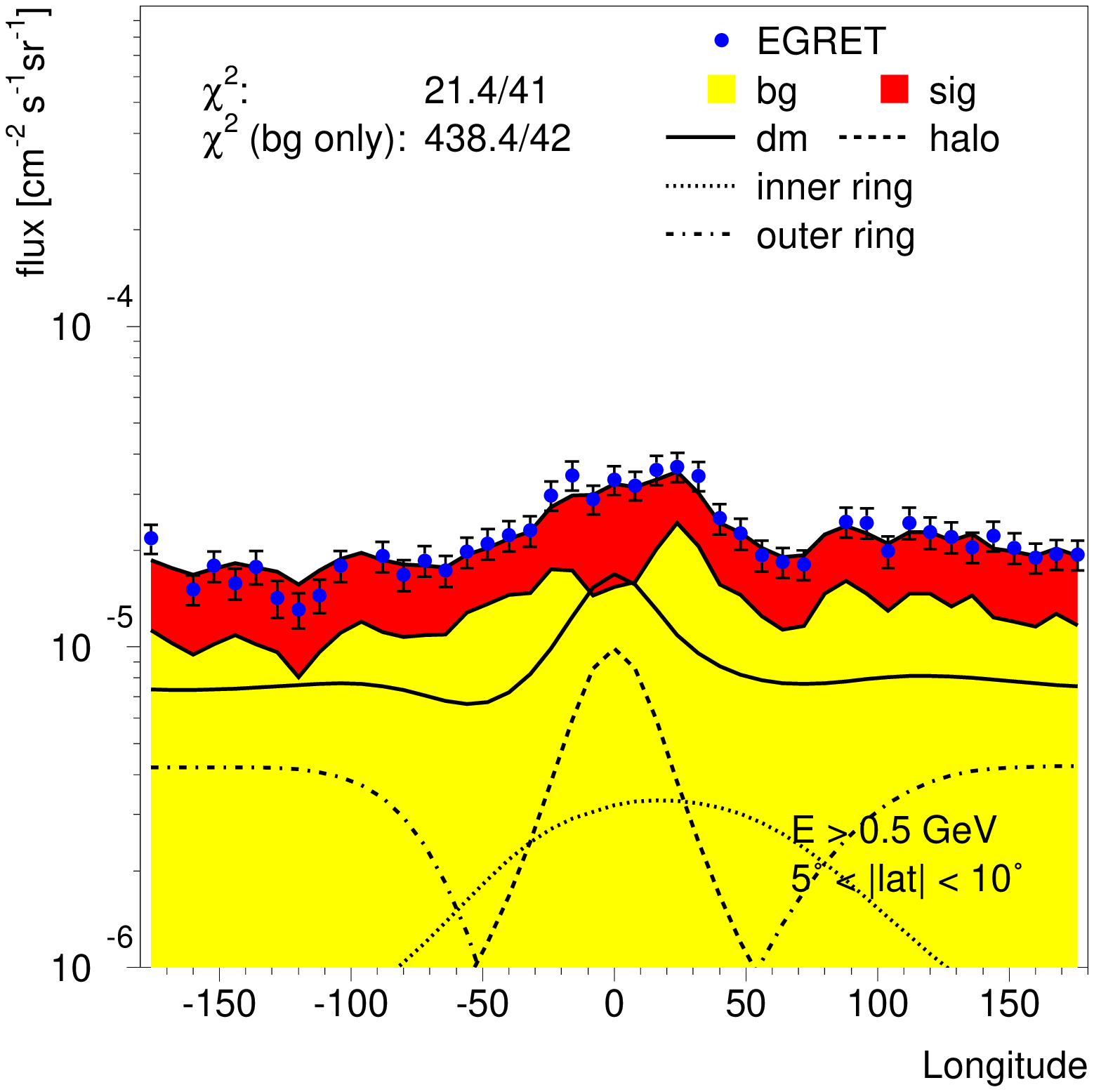}\\
  \includegraphics[width=0.3\textwidth,clip]{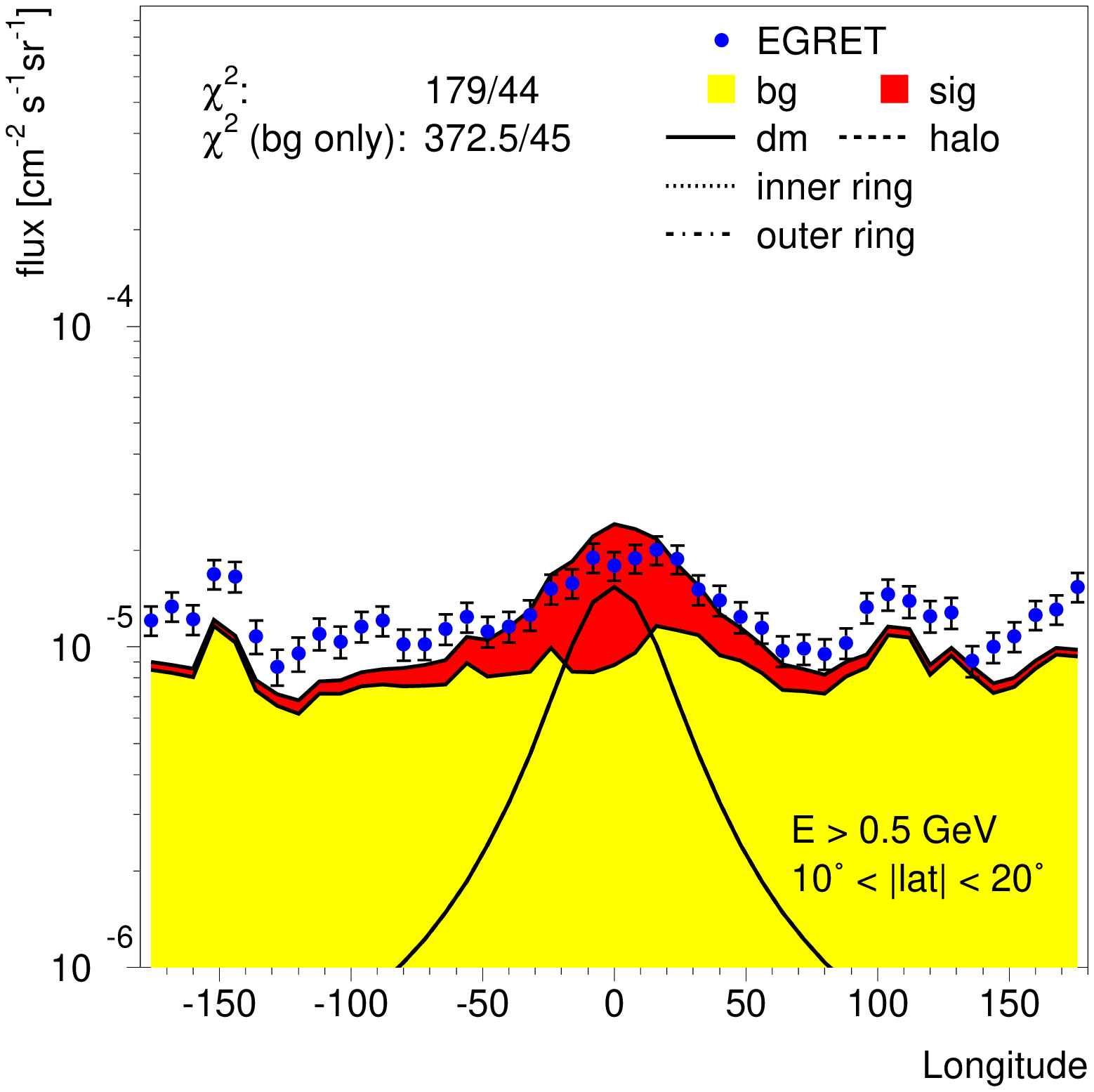}
 \includegraphics [width=0.3\textwidth,clip]{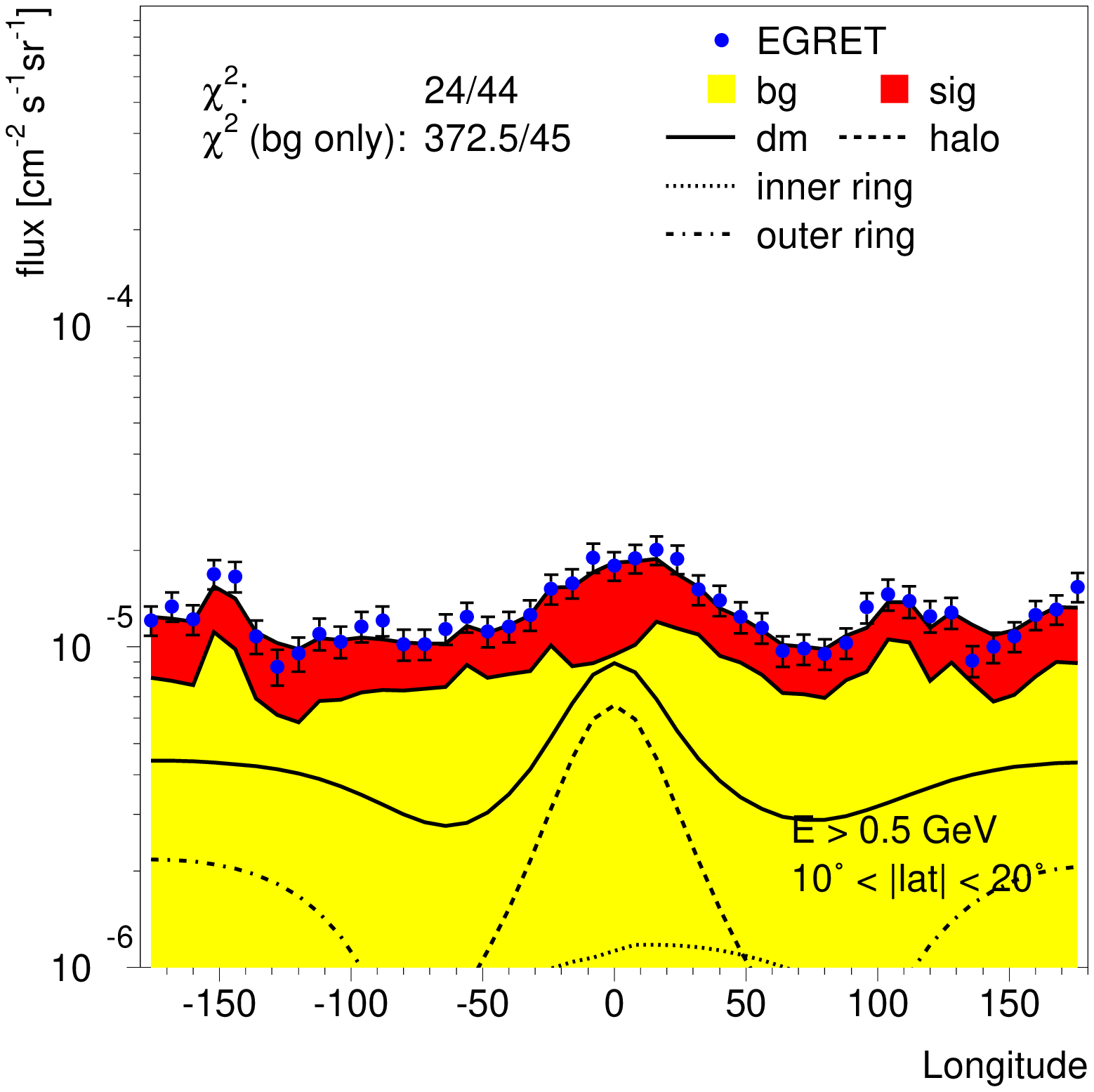}\\
 \includegraphics [width=0.3\textwidth,clip]{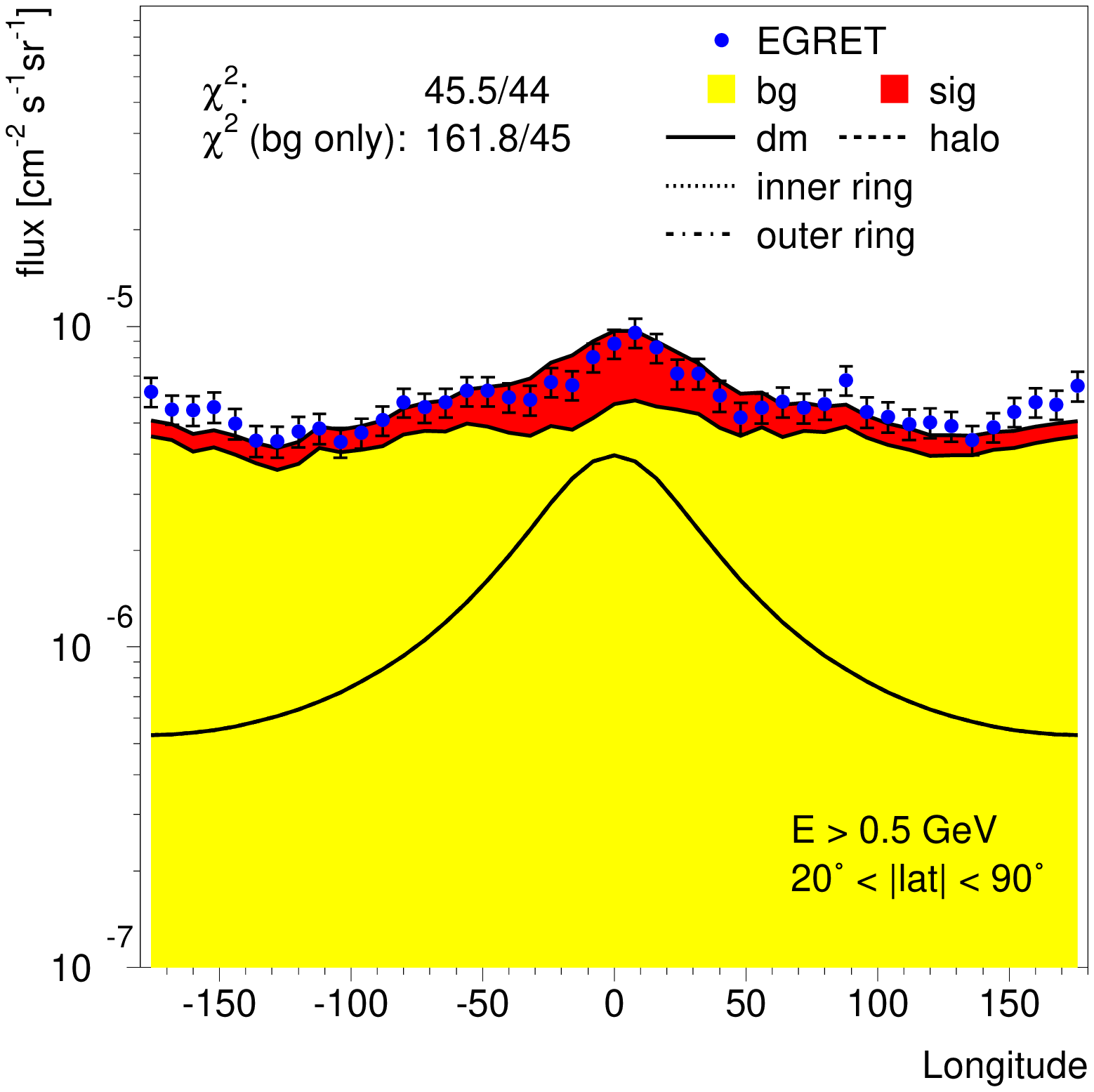}
 \includegraphics [width=0.3\textwidth,clip]{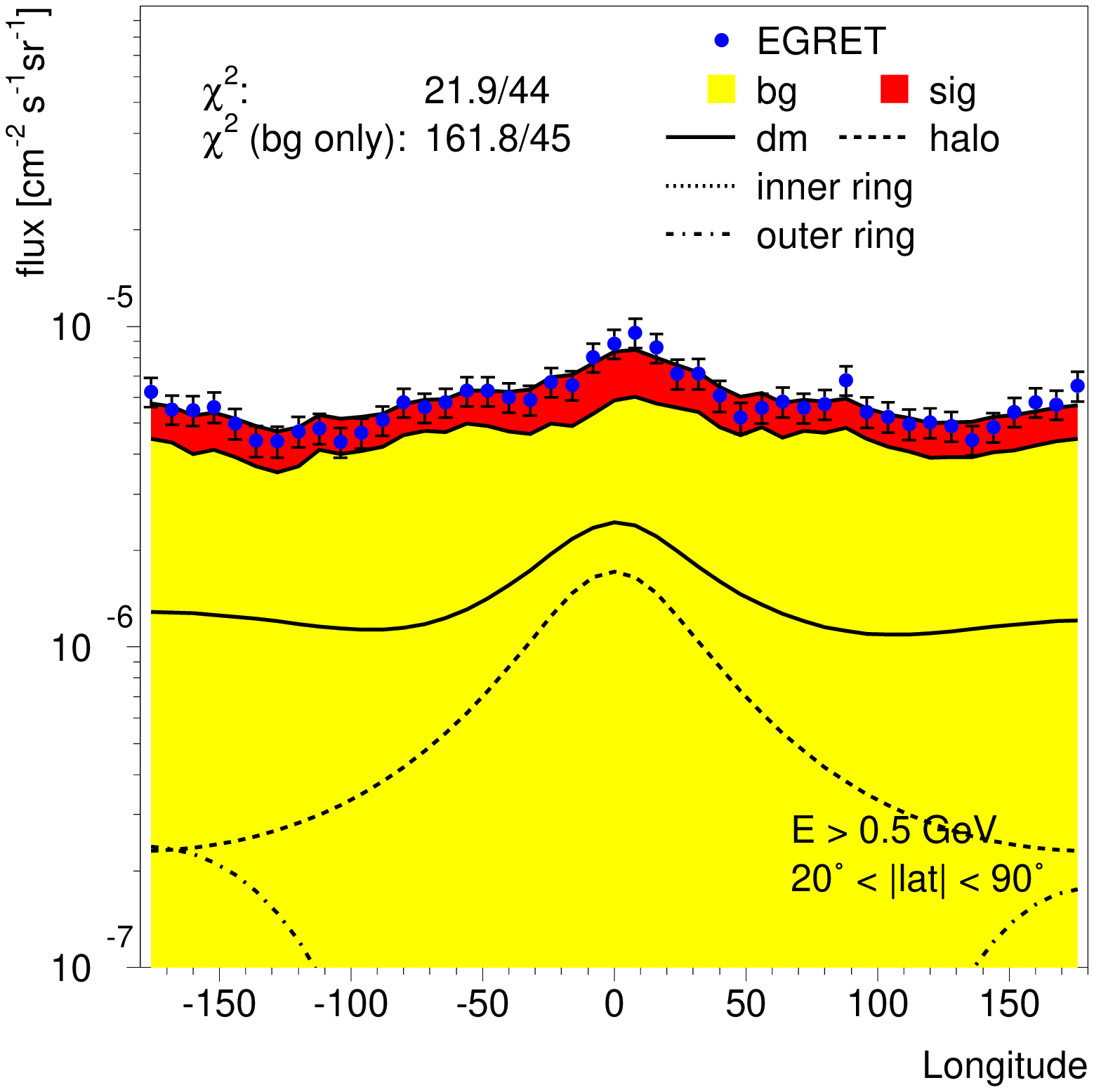}
 \caption[]{Top row:
 the longitude distribution of diffuse gamma-rays for  latitudes $0^\circ<|b|<5^\circ$
for the isothermal profile without (left) and with rings (right).
 The points represent the EGRET data.
 The contributions from the background and the neutralino annihilation signal have
  been indicated by the light (yellow) and dark (red) shaded area, respectively
  and the positions of the rings are indicated as well.
  The following panels: as above for latitudes $5^\circ<|b|<10^\circ$,
  $10^\circ<|b|<20^\circ$ and $20^\circ<|b|<90^\circ$.}
 \label{long}
\end{center}
\end{figure}
\begin{figure}[t]
\begin{center}
 \includegraphics [width=0.4\textwidth,clip]{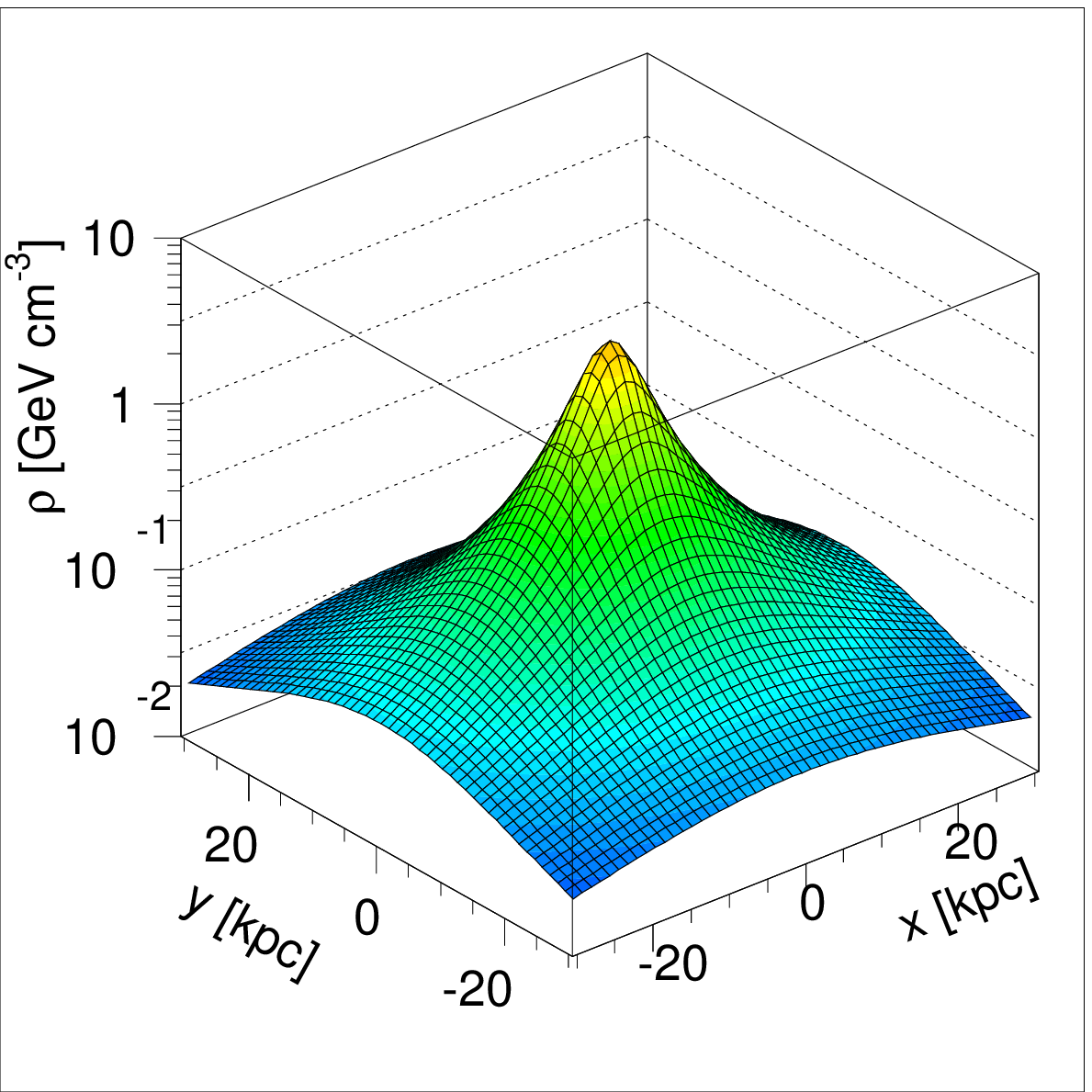}
 \includegraphics [width=0.4\textwidth,clip]{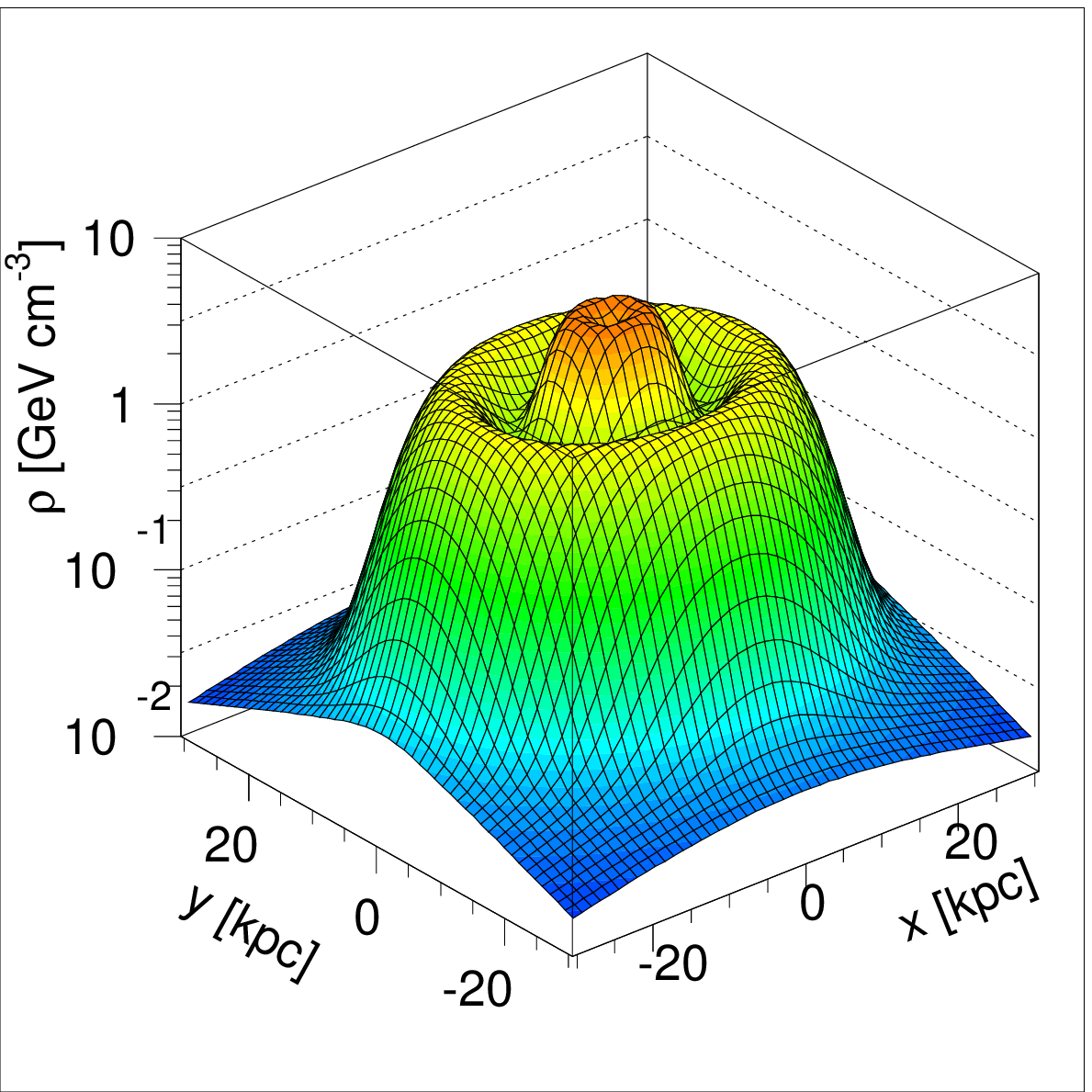}
 \includegraphics [width=0.4\textwidth,clip]{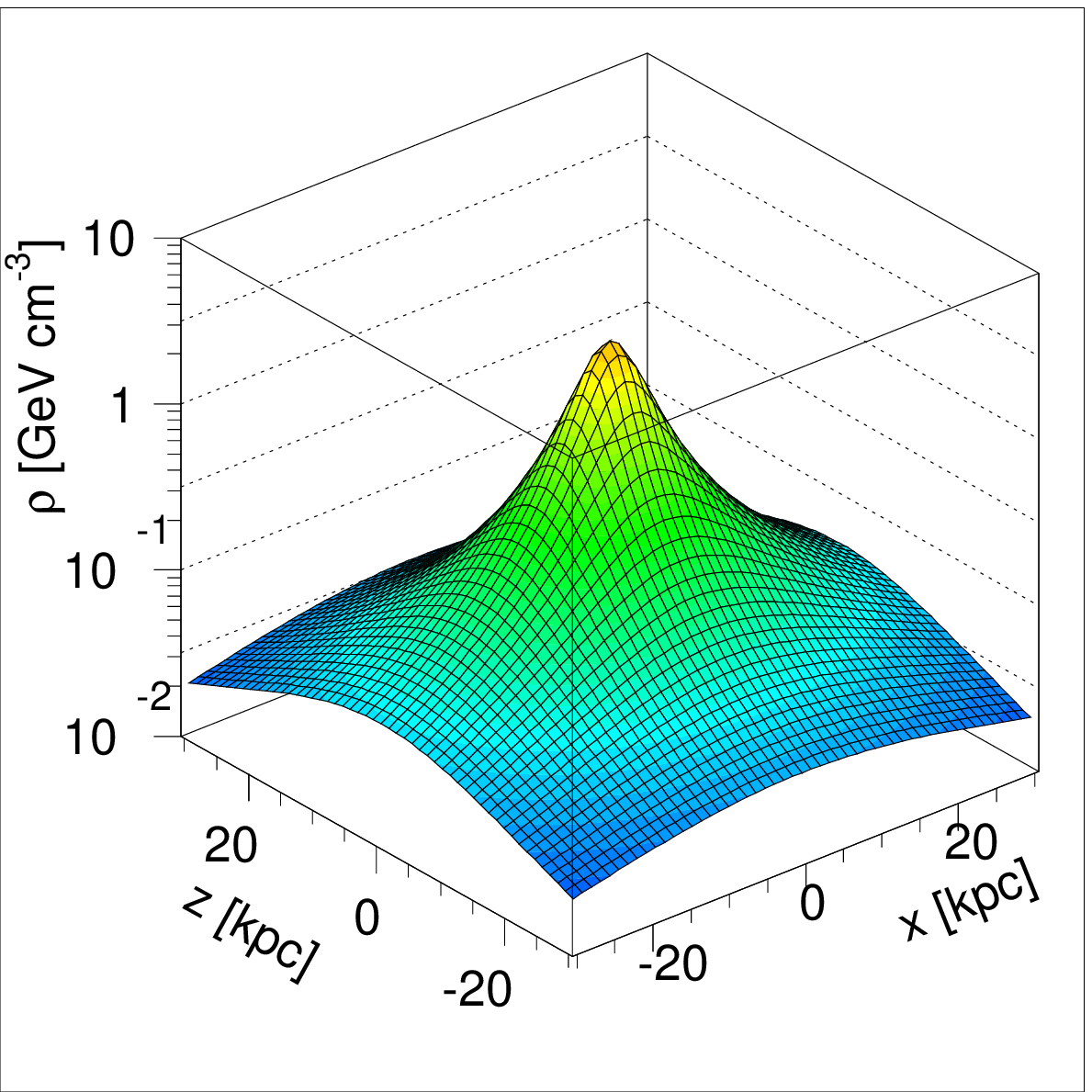}
 \includegraphics [width=0.4\textwidth,clip]{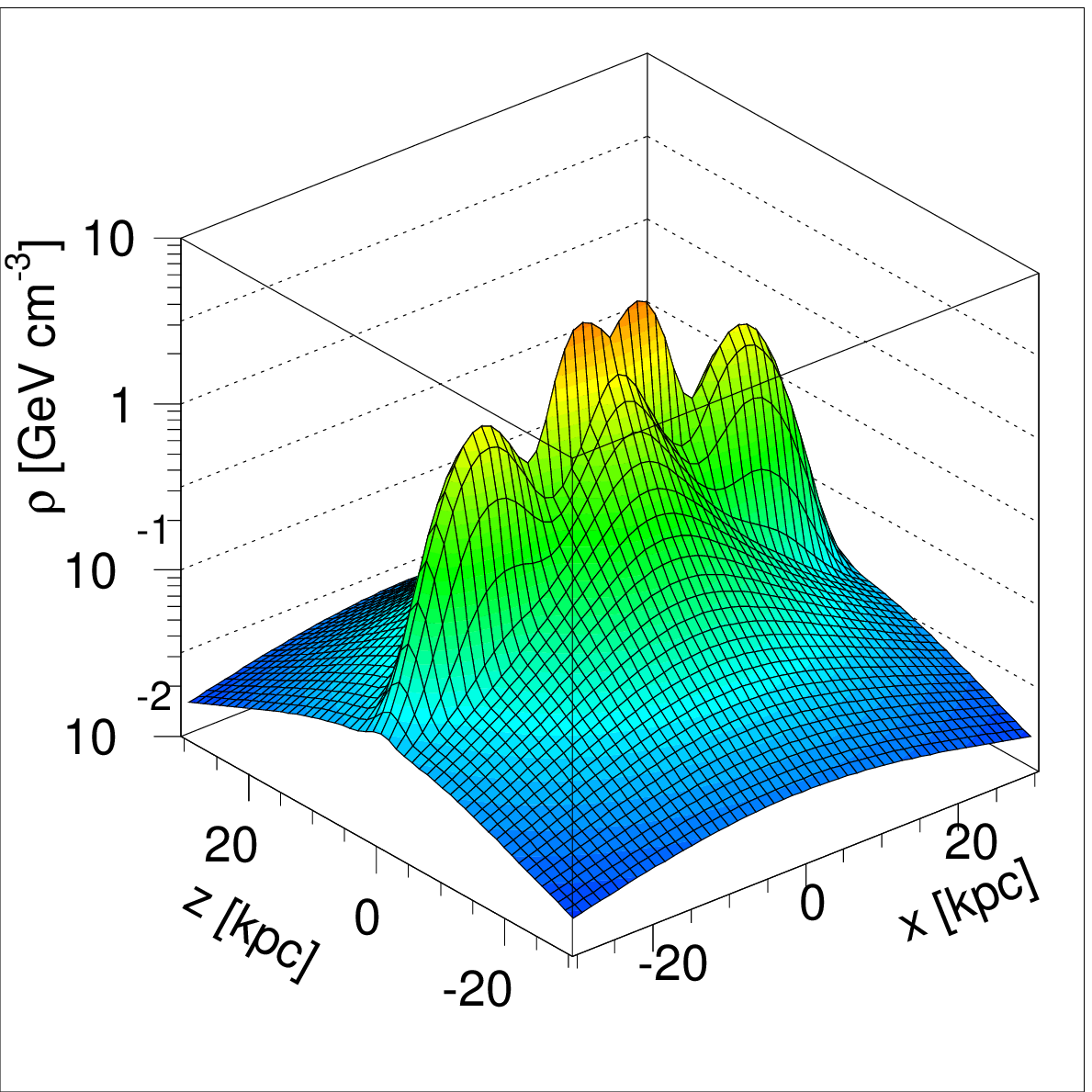}
 \caption[]{
  3D-distributions of the isothermal haloprofile in the xy- (top row) and xz-plane (bottom row)
  without (left) and with (right) rings. The elliptical shape (b/a=0.8,c/a=0.9) and
  ring structures in the disc (z=0 plane) are
  clearly seen.}
 \label{profile}
\end{center}
\end{figure}
\begin{figure}[t]
\begin{center}
 \includegraphics [width=0.4\textwidth,clip]{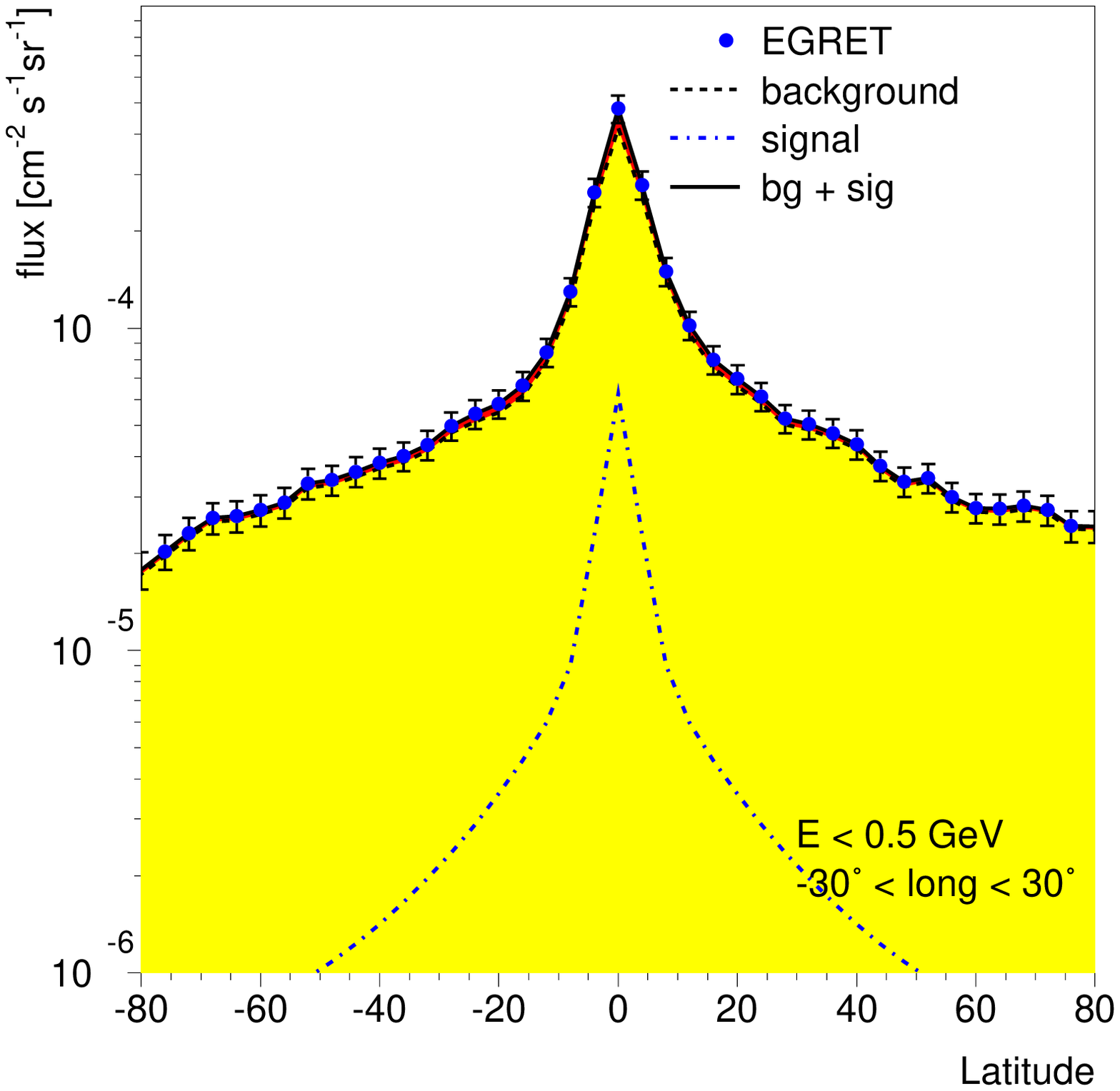}
 \includegraphics [width=0.4\textwidth,clip]{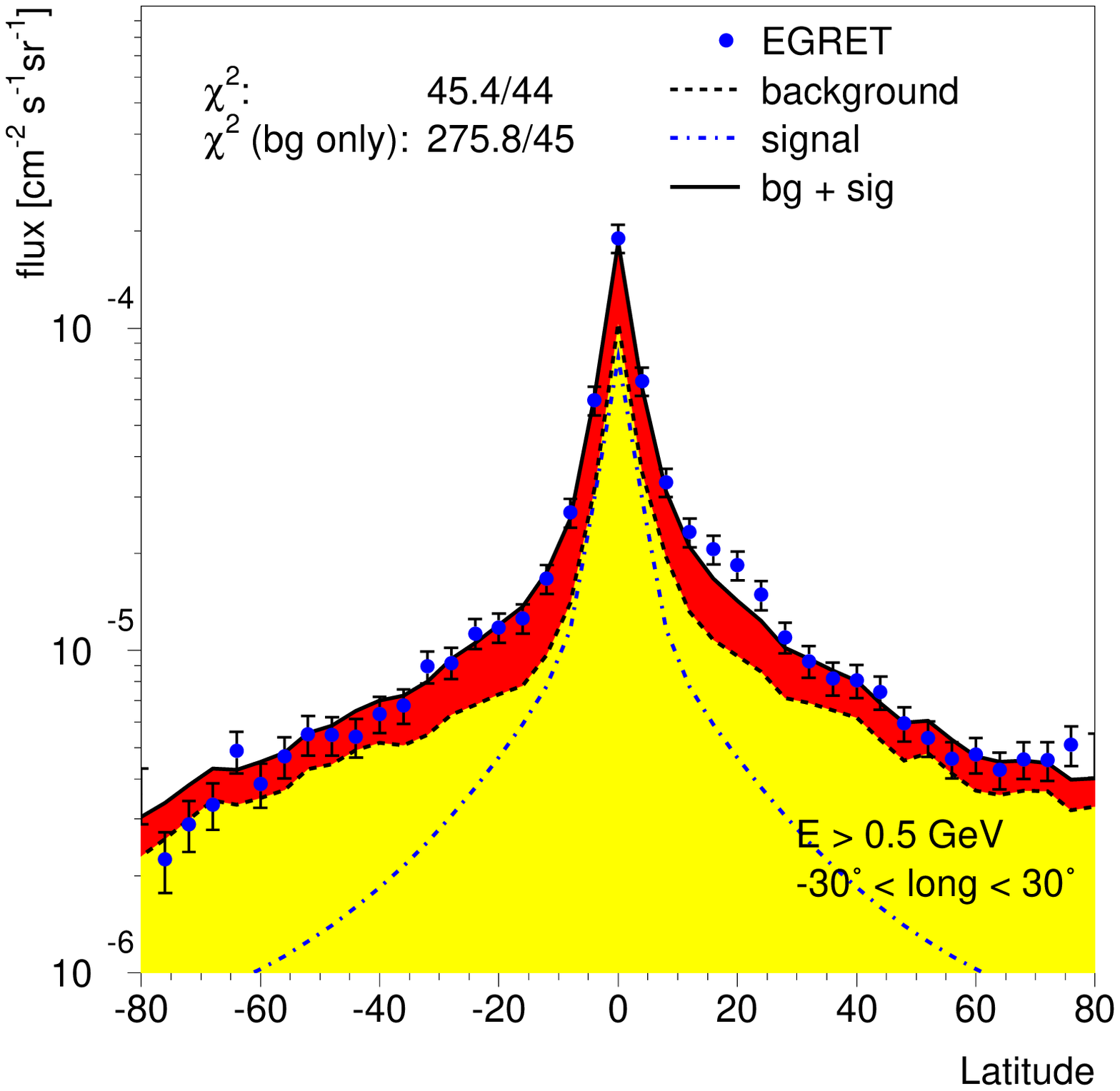}
 \caption[]{
 The latitude distribution of diffuse gamma-rays for  longitudes $-30^\circ<l<30^\circ$
 and two energy bins: $E_\gamma<0.5$ GeV (left) and $E_\gamma>0.5$ GeV (right).
 The points represent the EGRET data.
 The contributions from the background and the neutralino annihilation signal have
  been indicated by the light (yellow) and dark (red) shaded area, respectively.
  Note that the normalizations of the low energy data have been fitted for each point. The normalization
  factor is within the GALPROP uncertainty of 15\%.}
 \label{lat}
\end{center}
\end{figure}

\section{Determination of Halo Profile Parameters}\label{haloprofile}
The differential gamma flux in a direction forming an angle $\psi$ with the direction of
the galactic center is given by:
\begin{equation}
  \phi_\chi(E,\psi)=\frac{\langle \sigma v\rangle}{4\pi} \sum_f \frac{dN_f}{dE} b_f
  \int_{line~ of~ sight}B_l \frac{1}{2}\frac{\langle \rho_\chi^2\rangle}{M_\chi^2} dl_\psi
\label{gammafluxcont}
\end{equation}
where $b_f$ is the branching ratio into the tree-level annihilation final state, while
$dN_f/d E$ is the differential photon yield for the final state $f$. The WIMP mass density
along the line of sight,  $\rho_\chi$, enters critically in the prediction for the flux,
since the number of WIMP pairs is equal to $1/2\,\rho_\chi^2/M_\chi^2$. The factor $B_l$ is
the boostfactor, which represents the local enhancement of the number density with respect
to the average by the expected clustering of DM. Since the average of $\rho_\chi^2$ can be
significantly larger than $\langle \rho_\chi\rangle^2$  the boost factor can enhance the
flux  by one or two orders of magnitude. The cross section $\langle\sigma v\rangle$ is
taken from Eq. \ref{wmap} with the WMAP value for $\Omega h^2$.

 If one assumes that the clustering is similar in all directions, i.e. the same boostfactors
 every\-where, then the excess of diffuse gamma rays is just proportional to the square of
 the DM column density along the line of sight. The averaged density is
determined by the halo profile, which is normalized to the local DM density. The latter can
be estimated from the rotation curve of our galaxy to be $0.3~ GeV/cm^3$ for a spherical
profile and a distance between the sun and the galactic center $R_0=8.5$ kpc. For a
non-spherical profile this density has to be rescaled by integrating the total mass inside
$R_0$ and assuming $v_{rot}^2=G(\rho_{vis}+\rho_{dm})~dV/r=220 ~{\rm {km^2/s^2}}$.

 The halo parameters are fitted from the data in
the following way: the longitude distributions are determined in bins of 8$^\circ$ for four
different latitude ranges (0-5$^\circ$, 5-10$^\circ$, 10-20$^\circ$, 20-90$^\circ$), so one
has 4x45=180 angular bins. The data in each bin is split  in two energy ranges: data between 0.1 and 0.5 GeV and
data above 0.5 GeV, where only the high energy data has a significant contribution from DMA, as
can be seen from Fig. \ref{gamma_A}, so the low energy data can be used for the normalization of the background.
The normalization of the background was around 0.8 in
the plane of the galaxy and 1.1 for larger latitudes, indicating  small systematic
errors in the column densities.
After correcting for these systematic shifts, which are
within the quoted errors, the remaining systematic errors are about 10\%, as estimated from
fitting the energy spectrum e.g. at 0.1 and 0.5 GeV and determining the $\chi^2$ at 0.3
GeV. This was repeated for various energy bins and a $\chi^2/d.o.f.$  below one was
obtained for a systematic uncertainty of 10\%. This uncertainty will be used in all fits
and added in quadrature to the statistical error, which is usually much smaller.
Such a rescaling implies that we only rely on the shape of the background distribution,
as discussed before.

The fits to the 6 regions shown in Fig. \ref{gamma_us} were repeated for all the 180 intervals mentioned above.
The resulting  distributions are shown in Fig. \ref{long} for the energies above 0.5 GeV and an isothermal halo profile
 with and without rings. An isothermal profile without rings cannot describe the data, but
 adding two rings yields a good fit with a reduced $\chi^2=0.5$ for systematic errors of 10\%.
  The fitted
parameters are given in Table \ref{t2}. The density profiles with and without rings are
displayed in Fig. \ref{profile} and will be discussed  below. The boost factor for the profile
with rings is around 40, but can be changed easily by a factor two  up and down  
by either modifying the shape
of the background within errors and/or by changing the ellipticity and normalization of the halo profile.
Remember that the gamma flux is proportional to  $B_l\rho_\chi^2\propto B_l v_{rot}^4$, so
 increasing the contribution from the DM halo  to the rotation velocity by 20\% lowers the boost factor
already by a factor two.

 The assumption of a constant boost factor for all
directions is not necessarily true, since the tidal forces and merger history are a function of radius.
A radial dependence of the boostfactor would change the halo profile correspondingly.
With a constant boost factor and the fitted profile one obtains a reasonable $\chi^2$. Leaving the boost factors free
for the fitted  profile yields a boost factor variation of about 20\% for the 6 regions of Table \ref{t1},
indicating that  this procedure at least is self consistent.

 The NFW profile does not fit the data, neither with or without rings, since the longitudinal profile does
not show the strong cusp at the center. In order to be sure that the cusp is not subtracted
as a point source, the whole analysis was repeated by using the complete EGRET sky map,
i.e. the one including all the points sources. This does not change the results, since most
points sources  have a soft spectrum and even the many point sources near the center of the
galaxy increase the diffuse spectrum by less than 20\%. Three nearby sources  in the disc
(i.e. latitude $|b|< 5 ^\circ$) show a strong increase in the spectrum, namely at
longitudes of +85$^\circ$, -170$^\circ$ and -95$^\circ$. There seem to be several sources
in these regions, so the subtraction is not straight forward. Therefore  these 3 points
have been excluded in all analysis discussed before. The following sections discuss the
properties of the halo in more detail.
\subsection{Ring structure}
 Fig. \ref{halo}  displays the contributions from the
inner and outer rings at radii of 4.3 and 14 kpc, respectively.
 The maximum density  of the outer
ring is at a radius of 14 kpc with a one sigma spread  of 2.1 kpc in radius and 1.3 kpc
perpendicular to the plane. Its peak density is 2.3 GeV/cm$^3$, which corresponds to a
factor 8 enhancement over the density from the isothermal profile. These coordinates
coincide with the ring of stars observed in the plane of the galaxy at a distance of 14-18
kpc from the galactic center\cite{yanny,ibata}.  These stars show a much smaller velocity
dispersion (30 km/s) and larger z-distribution  than the thick disc, so it cannot be
considered an extension of the disc. A viable alternative is the infall of a dwarf
galaxy\cite{yanny,helmi},  for which one expects in addition to the visible stars a DM component.
From the size of the ring and its peak density one can estimate the amount of DM in the
outer ring to be approximately 8. 10$^{10}$ solar masses. Since the gamma ray excess covers
the full 360$^\circ$ of the sky, one can extrapolate the observed $100^\circ$ of visible
stars to obtain
 a total mass of  $\approx 10^9$ solar masses\cite{yanny}, so the
baryonic matter is only a small fraction of the total mass, which may explain
the stability and the small velocity
dispersion of the observed stars. The
outer ring of  the gamma rays excess shows an ellipticity with an 
axis ratio of about 0.8$\pm$0.1 with the major
axis close to the direction of the galactic anticenter.
The inner ring at 4.3 kpc with a width of 3.4 kpc in radius and  0.3 kpc in $z$ is more difficult to interpret,
since the density of the inner region is modified
by adiabatic compression and  interactions between the bar and the halo, as discussed before.
Furthermore, its parameters are strongly correlated with the parameters of the 
isothermal profile, so the sum of the two is an effective parametrization of 
the density near the centre.
The  ellipticity of the inner ring is quite strong (0.65$\pm$ 0.15) with its
major axis approximately in the direction of the bar, which suggests a strong influence of the
adiabatic compression by the bar. The enhancement of the density in 
the ring by a factor 2.8 over the density of the isothermal profile is 
the right order of magnitude\cite{adiabatic}.

It is interesting to note that the coordinates of the inner ring coincide with the ring of  
cold dense molecular hydrogen
gas, which reaches a maximum density at 4.5 kpc and has a width of 3 kpc as well (see e.g. Ref. \cite{book}).
This suggests that just like the ring of stars at 14 kpc is stabilized by the ring of DM,
this ring is stabilized by DM as well and 
the higher hydrogen density in the DM ring  increases the binding of the atoms into molecules.

\subsection{Halo ellipticity}
\begin{figure}[t]
\begin{center}
 \includegraphics [width=0.9\textwidth,clip]{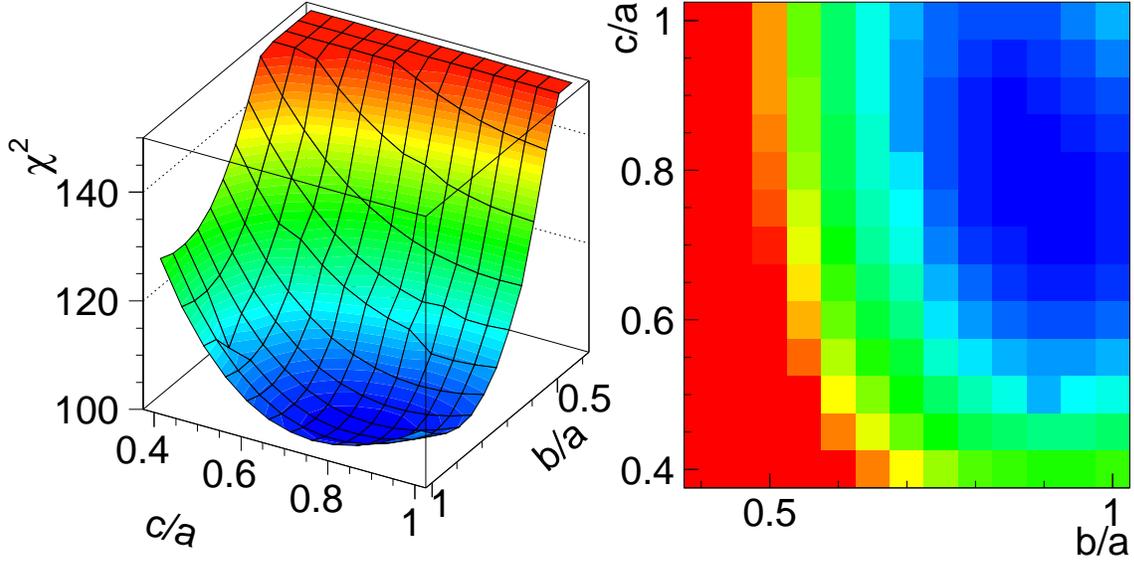}
 \caption[]{
  3D-distribution (left) of the $\chi^2$ fit to the 4 longitudinal distributions
  and its projection (right) as function of the ratio of the short to long axes.}
 \label{ellipticity}
\end{center}
\end{figure}

The  axis ratios of the basic halo profile turn out to be similar to the ellipticity of the
outer ring, i.e. $b/a$ and $c/a$  are around 0.8-0.9 in both directions.
 The errors of these ratios are about 0.1, as can be estimated from
the $\chi^2$ distribution shown in Fig. \ref{ellipticity}. For a  $c/a$ ratio of 0.8 the
most probably value from N-body simulations  for $b/a$ is 0.9\cite{Jing:2002np} in agreement
with the values found. 
The halo is prolate instead of oblate.  Oblate haloes are
 often assumed  in discussions on the rotation curves\cite{olling} or stability
of the ring of stars from the tidal disruption of  the Sagittarius 
dwarf\cite{Majewski:2003ec,Ibata:1996dv,martinez,Merrifield:2003hz}.
It would be interesting to study the stability of this ring in the proposed halo,
especially since the ring of stars is seen in a plane almost  perpendicular to the major 
axis of the prolate halo.
The alignment between the angular momentum and  the {\it minor} axis in a prolate halo
is a preferred structure, as  found 
in  recent high resolution N-body simulations by Bailin \& Steinmetz\cite{steinmetz}. 

\subsection{Comparison with rotation curve}\label{rotation}

\begin{table} [tbp]
 \begin{center}
  \begin{tabular}{|c|c|c|c|}
   \hline
   Parameter & Value&Parameter&Value \\
   \hline
   $\alpha$ & 2& $R_{a}$ & 4.3 kpc\\
   $\beta$ & 2&$\sigma_{R,a}$ & 2.6 kpc\\
   $\gamma$ & 0&$\sigma_{z,a}$ & 0.3 kpc\\
   $R_0$ & 8.5 kpc&$\rho_{b}$ & 1.9 GeV cm$^{-3}$\\
   $a$ & 4 kpc&$R_{b}$ & 14 kpc\\
   $\rho_0$ & 0.42 GeV cm$^{-3}$&$\sigma_{R,b}$ & 3.4 kpc\\
   $\rho_{a}$ & 3.4 GeV cm$^{-3}$&$\sigma_{z,b}$ & 1.7 kpc\\
   $b/a$&0.9&$c/a$&0.8\\
   \hline
  \end{tabular}
  \caption[]{Halo parameters for the isothermal profile with 2 rings.}\label{t2}
 \end{center}
\end{table}
\begin{figure}
\begin{center}
 \includegraphics [width=\textwidth,clip]{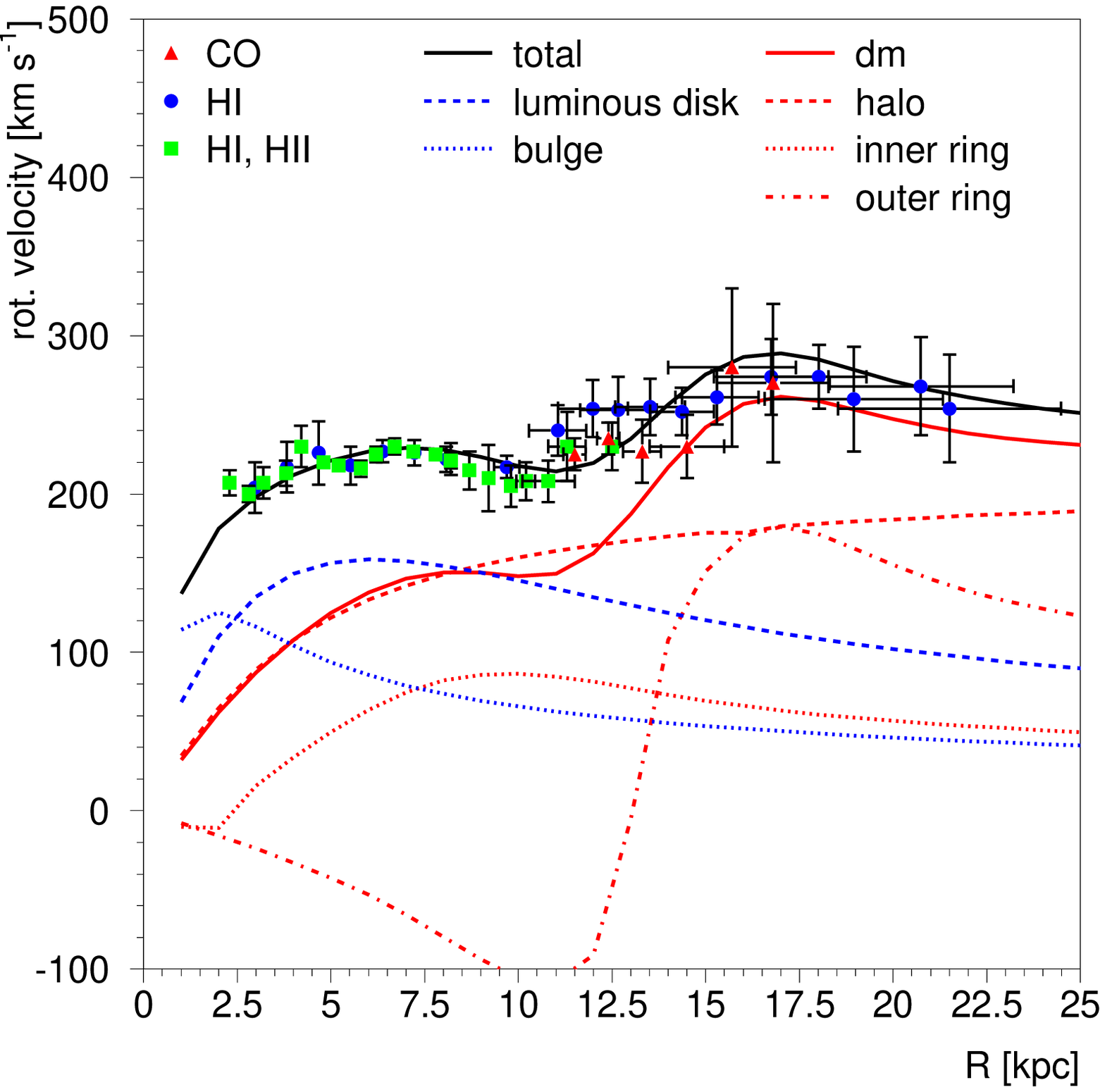}
 \caption[]{
 The rotation curve of the galaxy for the DM haloprofile of Fig. \ref{profile}.
 The data are from Refs. \cite{honma,brand,schneider}. The contributions from the individual masses
 have been indicated. Note the negative contribution of the massive gaussian ring of DM at 14 kpc,
 which exerts an outward and hence negative force  on a tracer well inside that ring.}
 \label{rot}
\end{center}
\end{figure}

The rotation curve of our galaxy shows a peculiar non-flat structure near our solar system,
namely at  $r=1.1R_0$ kpc the slope  changes
sign\cite{honma,brand,schneider,Binney:1996fb}. The two ring model describes this structure
 well, as shown in Fig. \ref{rot}. The contributions from each of the mass terms have been
 shown separately. The basic explanation for the negative contribution from the outer ring
 is that a tracer star at the
 inside of the ring at 14 kpc feels an outward force from the ring, thus a negative
 contribution to the rotation velocity. In order to calculate this more quantitatively
 one needs the complete  distribution of both the visible and DM mass.
The contributions from the baryonic matter to the rotation curve  were modeled as in Ref.
\cite{olling}, while the DM density was taken from Eq. \ref{halo1}.
The solution for the gravitational potential $\Phi$  of the Poisson equation can be written
 in spherical coordinates ($x=r\cos\phi \sin\theta,\ y=r\sin\phi \sin\theta,\ z=r\cos\theta $)
 as:
\begin{equation}
\Phi(r,\theta,\phi)=-\int_0^\infty r'^2dr'\int_{-1}^{1}d\cos\theta'
\int_0^{2\pi}d\phi'\frac{\rho(r',\theta',\phi')}
{\sqrt{r^2+r'^2-2rr'\sin\theta\sin\theta'\cos(\phi-\phi')-2rr'\cos\theta\cos\theta'}}
\label{phi1}
\end{equation}
 or in the plane in the direction $\phi=0, \theta=\pi/2$:
\begin{equation}
\Phi(r,\pi/2,0)=-\int_0^\infty r'^2dr'\int_{-1}^{1}d\cos\theta'
\int_0^{2\pi}d\phi'\frac{\rho(r',\theta',\phi')}{\sqrt{r^2+r'^2-2rr'\sin\theta'\cos(\phi-\phi')}}
\label{phi2}
\end{equation} Note that $\rho$ includes all masses. The rotation velocity for
a circular orbit at a radius $r$ can then be calculated by requiring that  the resulting
gravitational force on a tracer star  equals the centrifugal force, i.e. $v^2/r=F_G/m$ or
\begin{equation}
{v^2(r)\over r}=\frac{d\Phi(r)}{dr}=\int_0^\infty r'^2dr'\int_{-1}^{1}d\cos\theta'
\int_0^{2\pi}d\phi'\frac{\rho(r',\theta',\phi')(r-r'\sin\theta'\cos(\phi-\phi'))}
{(r^2+r'^2-2rr'\sin\theta'\cos(\phi-\phi'))^{3/2}}. \label{v2}
\end{equation}
This threefold integral was integrated numerically to obtain the contribution from all mass
elements in the halo.  The contributions of the bulge, halo and DM are shown separately in
Fig. \ref{rot}. The negative contributions from the rings originate from the fact that the
derivative of the gravitational potential $\Phi$ changes its sign, when crossing the
maximum of the ring and so does the contribution to $v^2$ (see term $r-r'$ in numerator of
Eq. \ref{v2}). This implies an outward gravitational force exerted by the ring for a tracer
inside the ring and an inward force for a tracer outside the ring\footnote{The
gravitational force is only zero in a spherical $shell$, not inside a $ring$.}. The
explanation for the hitherto mysterious change of sign of the slope near $r_c=1.1R_0$ finds
then its natural explanation in the large ring of DM at $r_c=14$ kpc, whose mass is
determined by the excess of energetic gamma rays.

\subsection{Comparison with the surface density}
The gravitational potential near our solar system is strongly constrained by the height
distribution $n(z)$ and velocity dispersion $\sigma_z$  of the stars in the disc. From
these measurements the surface density $\Sigma(<z)=\int_{-z}^{+z} \rho(z')dz'$ can be
determined for a given interval in $z$, which was found to be in the range 70-85 $M_\odot
pc^{-2}$\cite{kuijken,bienayme}. However, here the density includes all the mass. In order
to obtain the contributions of DM and baryonic matter separately one has to assume a height
distribution for each of them. For the visible matter density one usually assumes an
exponential decrease with height and for the DM a constant density, but the results still
vary widely depending e.g. on the value of scale height for the exponential decrease or the
DM density profile. A consensus value for the baryonic surface density is suggested by
Olling and Merrifield: $\Sigma_* =35\pm~ 10 M_\odot pc^{-2}$ \cite{olling}.
The assumption of a constant DM density is not valid in our halo model with rings in the
disc. Integrating the mass density distributions we find for the baryonic surface density
$\Sigma_{*} (|z|<1100) =30~ M_\odot pc^{-2}$ and for the DM surface density
$\Sigma_{DM}(|z|<1100) =60~M_\odot pc^{-2}$. The visible part is within the ``consensus''
range, but the total is above the value quoted by Kuijken and Gilmore\cite{kuijken},
however compatible with the more recent values from Siebert, Bienaym$\rm\acute e$ and
Soubiran\cite{bienayme}.  It should be noted that the surface density of the DM is rather
sensitive to the width of the inner ring, since the solar system is located on the falling
outer slope of this ring, so these numbers should be taken as indicative.

\subsection{Galactic parameters}
 From the DM halo profile and visible density, as derived from the rotation curve for the given
 DM profile one can determine the following basic properties of our galaxy:
1) The radius containing an average density 200 times the critical density equals
$R_{200}=295 ~kpc$ 2) The total DM mass inside this radius  is 3.0$\cdot10^{12}$ solar
masses to be compared with a visible mass of 5.5$\cdot10^{10}$ $M_\odot$ 3) The inner
(outer) ring contribute 0.6 (2.7) \% to the total DM mass 4)  The fraction
$f_d=M_{disc}/M_{DM}=0.018$
 5) The concentration parameter $c=R_{200}/r_c=295/4=74$. All these parameters are well inside the range expected from
N-body simulations for an isothermal profile\cite{Jimenez:2002vy} and independent mass
estimates of the galaxy\cite{wilkinson}.
\section{Is the WIMP signal compatible with Supersymmetry?}
%
\begin{figure}[t]
\begin{center}
 \includegraphics [width=0.45\textwidth,clip]{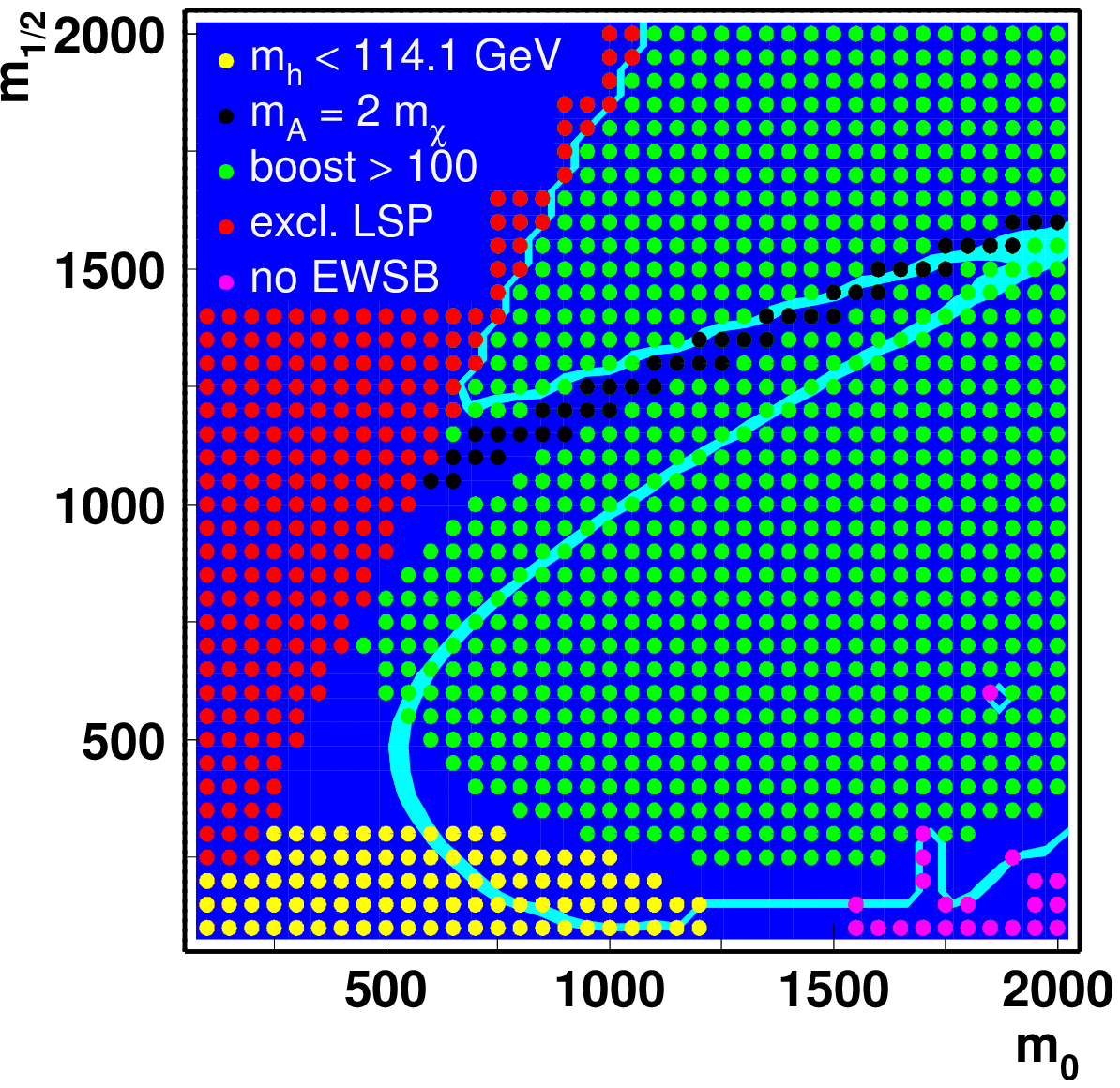}
 \includegraphics [width=0.45\textwidth,clip]{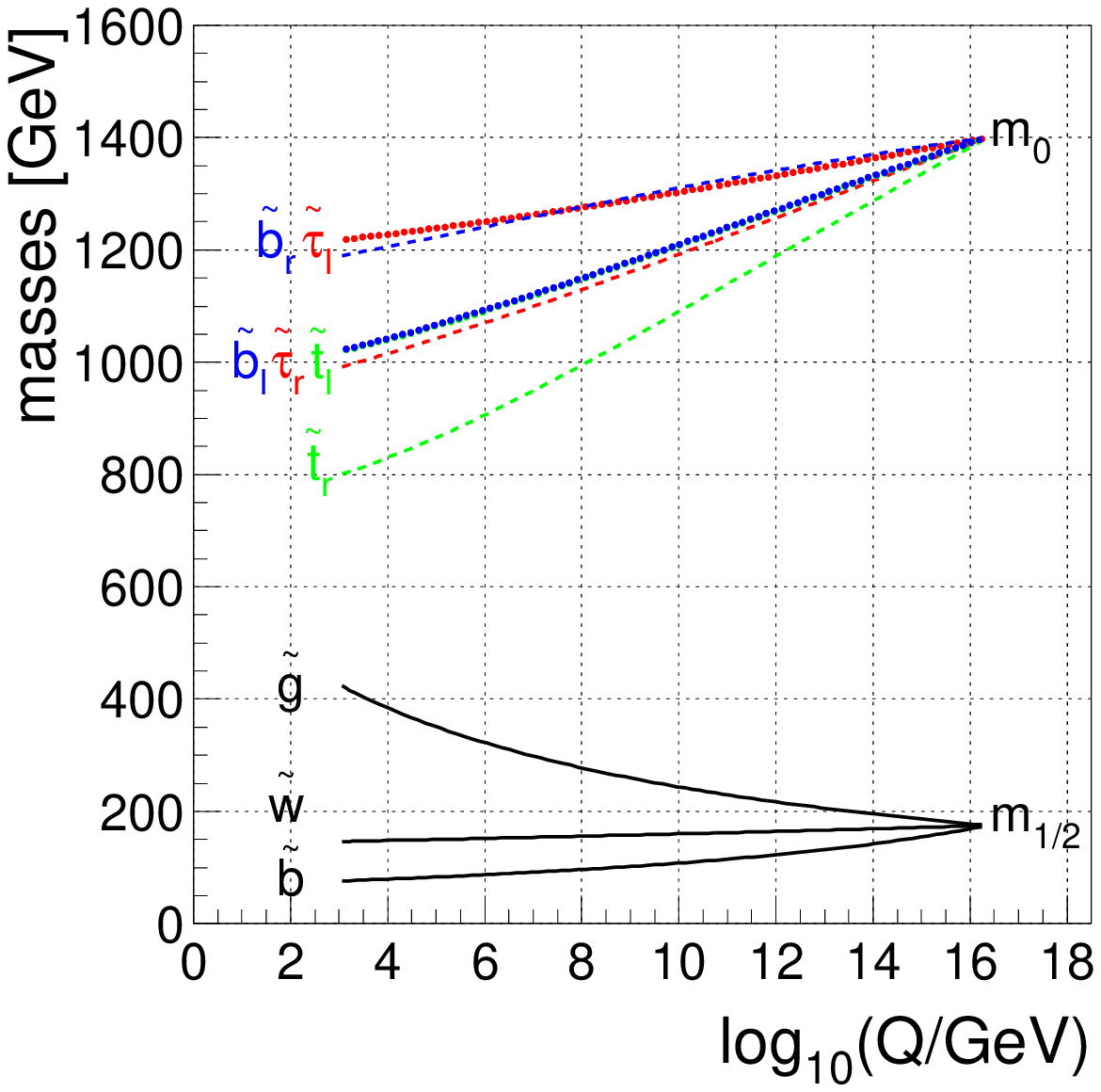}
 \caption[]{
 The light shaded (blue) line in the region allowed by WMAP in the $\mzero,\mhalf$ plane
for $\tb=51$, $\mu>0$ and $A_0=0.5\mzero$.
  The excluded regions, where the stau would be the LSP or EWSB fails or
 are indicated by the dots. Also the regions where the boost factor would be above 100 and
  the resonance region,
 where $|m_A-2m_{\chi_0}| \le 10$ GeV, are indicated.
The region for $\mhalf\approx 180$ and $\mzero\approx 1400$ all constraints from EGRET, WMAP and electroweak
data are fulfilled. The evolution of the  particle spectrum is shown on the right hand side, showing
that the squarks and sleptons have masses in the TeV range, while the gluinos and charginos are relatively light.
 }
 \label{relic}
\end{center}
\end{figure}

Supersymmetry~\cite{susyrev} presupposes a symmetry between fermions
and bosons, which can be realized in nature only if one assumes each particle
with spin j has a supersymmetric partner with spin $\vert j-1/2\vert$
($\vert j-1/2\vert$ for the Higgs bosons).
This leads to a doubling of the particle spectrum.
Obviously SUSY cannot be an exact symmetry of nature; or else the
supersymmetric partners would have the same mass as the normal
particles. The mSUGRA model, i.e. the Minimal Supersymmetric
Standard Model (MSSM) with supergravity inspired breaking terms,
is characterized by only 5 parameters: $m_0,~m_{1/2},~\tb,~\mbox{sign}(\mu),
~A_0$. Here $m_0$ and $m_{1/2}$ are the common masses for the
gauginos and scalars at the GUT scale, which is determined by the
unification of the gauge couplings. Gauge unification is still
possible with the precisely measured couplings at LEP~\cite{bs}.
The ratio of the vacuum expectation values of the two Higgs
doublets is called \tb ~ and $A_0$ is the trilinear coupling at
the GUT scale. We only consider the dominant trilinear couplings
of the third generation of quarks and leptons and assume also
$A_0$ to be unified at the GUT scale.
The absolute value of the Higgs mixing parameter $\mu$ is
determined by electroweak symmetry breaking, while its sign is
taken to be positive, as preferred by the anomalous magnetic
moment of the muon~\cite{bs}.

The Lightest Supersymmetric Particle (LSP) in the mSUGRA scenario is one of the four
neutralinos, which are mixtures of the four spin 1/2 partners of the photon, Z-boson and
two Higgs particles. If R-parity is conserved the LSP is stable. In a large region of
parameter space it has a large gaugino component, so the couplings to other particles are
similar to  the photon coupling, but R-parity conservation forbids electromagnetic
couplings to normal matter. Therefore the LSP can have only weak couplings, thus being a
perfect candidate for DM.  The neutralino annihilation cross section depends on the SUSY
parameters. For the neutralino annihilation cross section from the precise relic density
from the WMAP data only a small range of SUSY masses is allowed, as shown by the thin line
in Fig. \ref{relic} for $\tb=51$, $\mu>0$ and $A_0=0.5\mzero$; the latter values turn out
to be good parameters if one requires relatively low boost factors for the mass ranges
preferred by WMAP and EGRET simultaneously. The relic density has been calculated as
function of the SUSY parameters with the program microMegas\cite{micro} after interfacing it
to the Suspect program\cite{suspect} to calculate the SUSY mass spectrum  from the
Renormalization Group Equations (see e.g. \cite{susyrev}) and to the FeynHiggs
program\cite{fhf} to calculate the Higgs mass in Supersymmetry. Several regions are
excluded in this plane. The corner at the bottom left is excluded because the Higgs mass is
predicted to be below the experimental lower limit of 114.4 GeV, as measured at
LEP\cite{lep}. In the corner at the bottom right the radiative corrections from the
gauginos, which are light there, are too small to induce electroweak symmetry breaking. In
the corner at the top left the Lightest Supersymmetric Particle (LSP) would be charged, as
expected, since here the GUT scale values of the neutralinos are high (large $\mhalf$) and
the squarks and sleptons are light (small $\mzero$). Since the DM is known to be neutral,
this region has to be excluded.  In the adjacent region  the stau cannot decay fast into a
neutralino and tau (because the mass difference between neutralino and stau is less than
the tau mass), but a stau and neutralino can annihilate into a tau plus photon. This
coannihilation reduces the relic density to values required by the WMAP data, but these
regions require large boost factors, since in the present galaxy the next heavier LSP's
(NLSP) have decayed and only the self annihilation contributes. The regions, where the
boost factors are less than 100 are shown in Fig. \ref{relic} as well.  For boost factors
below 100 only two regions are allowed by the WMAP data (blue line): one around
$\mzero=600, ~\mhalf=400$ GeV and one around $\mzero=1400, ~\mhalf=180$ GeV.
 The latter region  is strongly preferred by the EGRET spectrum.
 The evolution of the sparticle masses from the GUT scale values towards lower energies
 is shown in  the right panel of Fig. \ref{relic}.
 The large value of $\mzero$ yields squark and slepton masses around 1 TeV, but
the gluinos and charginos are relatively light, which would have interesting consequences
for searches at future colliders. The compatibility with Supersymmetry implies the
possibility that the Dark Matter is the supersymmetric partner of the Cosmic Microwave
Background. Future experiments from 2007 onwards at the Large Hadron Collider under
construction at CERN will tell, since the predicted sparticle masses from the present
analysis are  within the range of this accelerator with a centre of mass energy of 14000
GeV.

\section{Summary}

In this analysis the diffuse gamma rays from  dark matter annihilation and background are
separated by the difference in the shape of the energy spectra,  thus eliminating the need
to rely on  absolute fluxes. The normalization of the background comes out to be close to
the absolute prediction of the GALPROP conventional propagation model of our galaxy, while
the normalization of the DM signal corresponds to a boost factor from 20 upwards, if the
annihilation cross section is taken from WMAP data.  Larger boost factors are always
possible, if in addition to self annihilation the coannihilation between WIMP and other
particles is allowed.

It is shown that the gamma ray excess  shows features expected from Dark Matter annihilation:
\begin{itemize}
\item The excess shows the same spectrum in all sky directions \item The intensity of the
excess is compatible with a prolate triaxial halo with the major axis in the plane of the
galaxy close to  the direction of the galactic anticenter and axis ratios  $b/a\approx
0.8\pm0.1$ and $ c/a \approx 0.9\pm0.1$. \item There is  additional excess in the galactic
plane located at two radii. The excess at r=14 kpc coincides with the ghostly ring of
stars, thought to originate from the tidal disruption of a dwarf galaxy. Since such a dwarf
galaxy usually carries a large amount of DM, the excess of the diffuse gamma rays in this
region is naturally explained by DMA. The inner ring at a radius of 4.3 kpc coincides with
the ring of high density molecular hydrogen gas, so its existence and stability can be
explained by the potential well from the DM. The higher density of atomic gas in such a
well will increase the binding into molecules. The inner ring has a small axis ratio
of 0.66 with the long axis in the direction of the bar, which suggests that
this enhancement is connected with adiabatic compression and/or additional
remnants of the merger history. The enhancement of the inner ring density 
over the isothermal halo is 2.7, which
is the right order of magnitude for adiabatic compression. \item Additional evidence that
the ring of stars is associated with a large amount of DM is provided by the rotation
curve: the hitherto mysterious change of slope  at a radius of 1.1$R_0$ finds its natural
explanation in the ring of DM, whose mass is calculated from the excess of gamma rays
 to be $8\cdot10^{10}~M_\odot$, which is two orders of magnitude larger
  than the estimated visible mass.
 \item The local surface density
 is  dominated by Dark Matter in our model, but           compatible with  the local
gravitational potential, as determined from the height distribution
and velocity dispersion of nearby older stars.
\item
The most probable WIMP mass is determined
from the spectrum of the excess to be in the range 50-100 GeV,
assuming the gamma rays in DMA originate from  $\pi_0$ decays.
\end{itemize}

Alternative models trying to explain the EGRET excess have to assume that
the locally measured fluxes of protons and electrons are not representative for
our galaxy, in which case these spectra  outside our local bubble can be tuned to 
obtain the more energetic gamma rays needed for the EGRET excess\cite{strong_egret}.
Of course such models cannot explain simultaneously the stability of the ring of stars
at 14 kpc and the change of slope in the rotation curve at $r=1.1R_0$.

The results mentioned above make no assumption on the nature of the Dark Matter, except
that its annihilation produces hard gamma rays  with a spectrum given by the EGRET excess.
A WIMP mass of 50-100 GeV is consistent with the Lightest Supersymmetric Particle predicted
in the Minimal Supersymmetric Model with supergravity inspired symmetry breaking.
 Scalar masses of squarks and sleptons are expected to be in the TeV range
for the parameters consistent with the EGRET,  WMAP and electroweak data.

It should be emphasized that  the excess of diffuse gamma rays has a statistical
significance of at least 10 $\sigma$ if fitted to the shape of the diffuse gamma ray
background obtained from the  conventional propagation model of our galaxy using the
GALPROP code. This combined with  all  features mentioned above provides an intriguing hint
that this excess is indeed  a signal from Dark Matter Annihilation.

\section{Acknowledgements}
We thank V. Moskalenko and A. Strong for sha\-ring with us all their knowledge about our
Galaxy and help in running their GALPROP code, O. Reimer for numerous discussions on EGRET data and 
to provide us with the unpublished data  up to 120 GeV
and D. Aubert, L. Bergstr$\rm\ddot o$m,  O.
Bienaym$\rm\acute e$, R. Ibata, A. Phukov, J. Primack, P. Sikivie and V. Springel for
useful discussions.
 This work was supported by the DLR
(Deutsches Zentrum f\"ur Luft- und Raumfahrt) and a
grant from the DFG (Deutsche Forschungsgemeinschaft, Grant 436 RUS 113/626/0-1).

\end{document}